\documentclass[hidelinks,onefignum,onetabnum]{siamart220329}
\usepackage{lipsum}
\usepackage{amsfonts}
\usepackage{graphicx}
\usepackage{epstopdf,epsfig}

\usepackage{algorithmic}
\ifpdf
  \DeclareGraphicsExtensions{.eps,.pdf,.png,.jpg}
\else
  \DeclareGraphicsExtensions{.eps}
\fi

\newsiamremark{remark}{Remark}
\newsiamremark{hypothesis}{Hypothesis}
\crefname{hypothesis}{Hypothesis}{Hypotheses}
\newsiamthm{claim}{Claim}
\headers{Multi-resolution simulations of ions}{J. Zhang and R. Erban}

\title{From Brownian dynamics to Poisson-Nernst-Planck equations: multi-resolution simulations of ions\thanks{Submitted to the editors DATE.\funding{
This work was supported by the Engineering and Physical Sciences Research Council, grant
number EP/V047469/1, awarded to Radek Erban. Jinyuan Zhang would like to thank Guangdong Postdoctoral Research Institute for studentship funding.}}}

\author{
Jinyuan Zhang
$\quad$\and$\quad$
Radek Erban$\,$\thanks{Mathematical Institute, University of Oxford,
Andrew Wiles Building, Radcliffe Observatory Quarter, Woodstock 
Road, Oxford OX2 6GG, United Kingdom}
}

\usepackage{amsopn}

\begin{document}

\maketitle

\begin{abstract} \noindent
Starting with a microscopic (individual-based) Brownian dynamics model of charged particles (ions), its macroscopic description is derived as a system of partial differential equations that govern the evolution of ion concentrations in space and time. The macroscopic equations are obtained in the form of the Poisson-Nernst-Planck system. A multi-resolution method for simulating charged particles is then developed, combining the detailed Brownian dynamics model in a part of the computational domain with coarser macroscopic equations in the remainder. The strengths, limitations, and applicability of microscopic, macroscopic, and multi-resolution simulation approaches are demonstrated through an illustrative model comprising a system of Na$^+$ and Cl$^-$ ions. 
\end{abstract}



\section{Introduction}
At the molecular level, different classes of simulation methods can be applied to reaction-diffusion-advection processes in biological systems, including particle-based (individual-based) methods, which treat each particle separately, and coarse-grained methods, which describe particle concentrations~\cite{Erban:2020:SMR}. Common coarse-grained approaches are written in terms of mean-field partial differential equations (PDEs), for example, the Poisson-Nernst-Planck (PNP) system, which is critical for understanding the dynamics of charged particles. The PNP system is particularly significant when examining the interactions between bulk and surface fixed charge~\cite{Bazant:2004:DCD,Klika:2023:UPN}. In biology, the PNP system has been used for analyzing ionic currents through protein channels embedded in  membranes~\cite{Eisenberg:1998:ICB,Kurnikova:1999:LRA,Cardenas:2000:TPN,Hollerbach:2001:TTP}. 

There are two state variables in the PNP system, the charge density of different ionic species and the electrostatic potential. In Section~\ref{sec2}, we will consider the system of two ionic species consisting of positive and negative ions in domain $\Omega \subset {\mathbb R}^3$. Denoting by $c^+: \Omega \to [0,\infty)$ and $c^-:\Omega \to [0,\infty)$ the concentrations of the positive and negative ions, respectively, we obtain the Nernst-Planck equation by combining diffusive flux resulting from a concentration gradient with the flux arising from a potential gradient in the form
\begin{equation}
\frac{\partial c^\pm}{\partial t}
=
\nabla \cdot D^\pm
\!\left( 
\nabla c^\pm
+
\frac{q^\pm}{k_b T} \, c^\pm \, \nabla \phi
\right),
\label{NernstPlanck}
\end{equation}
where $D^+$ (resp. $D^-$) is the diffusion constant of the positive (resp. negative) ion, $\phi$~is the electrostatic potential, $q^\pm$ is the charge, $k_b$ is the Boltzmann constant and $T$ is the absolute temperature. To get the PNP system, we couple equation~(\ref{NernstPlanck}) with the Poisson equation for the electrostatic potential
\begin{equation}
\nabla^2 \phi
= -\frac{1}{\varepsilon_0 \, \varepsilon} \left[\varrho_{\mathrm{p}} 
\,+\, 
q^+ \, c^+ 
\,+\, 
q^- c^-\right],
\label{Poisson}
\end{equation}
where $\varepsilon$ is the dielectric permittivity constant of the medium, $\varepsilon_0$ is the dielectric permittivity constant of the vacuum and $\varrho_{\mathrm{p}}$ is the permanent charge density.

The PNP system~(\ref{NernstPlanck})--(\ref{Poisson}) provides a mean-field description of a system of interacting charged particles. A more detailed description can be written using particle-based methods, such as molecular dynamics and Brownian dynamics (BD) simulations~\cite{Erban:2020:SMR,Erban:2016:CAM}. Considering the system of $N$ positive and $N$ negative ions, and denoting their positions at time $t$ by
\begin{equation}
\mathbf{X}_j^\pm(t)=[X_{j,1}^\pm(t),X_{j,2}^\pm(t), X_{j,3}^\pm(t)] \; \in \; \Omega \,,
\qquad \mbox{for} \quad j=1,2,\dots,N \,,
\label{ionpositions}
\end{equation}
where $\mathbf{X}_j^+$ (resp. $\mathbf{X}_j^-$) is the position of the $j$-th positive (resp. negative) ion, then the simplest BD model is given by the overdamped Langevin dynamics
\begin{equation}
\mbox{d} \mathbf{X}_j^\pm(t)
= 
\sqrt{2D^{\pm}} \,
\mbox{d} \mathbf{W}_j,
\qquad \mbox{for} \quad
j=1,2,\dots,N,
\label{eq0}
\end{equation}
where $\mathbf{W}_j$ is the vector of three independent Wiener processes~\cite{Erban:2020:SMR}. In computational studies, the BD equation~(\ref{eq0}) is commonly either discretized using fixed time step~$\Delta t$~\cite{Andrews:2010:DSC,Robinson:2015:MRD}, or solved by using event-based algorithms, such as Green's function reaction dynamics~\cite{Takahashi:2010:STC}, where the state of the system is updated by calculating the time at which the next interaction of particles occurs. The BD equation~(\ref{eq0}) provides a macroscopic description of more detailed molecular dynamics, Langevin dynamics or stochastic coarse-grained models~\cite{Erban:2014:MDB,Erban:2020:CGM}.

Although BD modelling based on equation~(\ref{eq0}) has previously been used for simulating ions in applications to calcium signalling~\cite{Flegg:2013:DSN,Dobramysl:2016:PMM}, the BD equation~(\ref{eq0}) must be generalized for use in other contexts. Considering applications to the dynamics of ions and ion channels~\cite{Corry:2000:TCT,Chen:2014:BDM,Song:2011:TAN,Im:2002:IPS}, we generalize equation~(\ref{eq0}) by adding a drift term which includes a gradient of the (electric) potential. The resulting BD model is formulated in Section~\ref{sec2} as equation~(\ref{BDSDEform}). This additional drift term depends not only on the position of the ion, but also on positions of other ions and on an external (electrical) field. In Section~\ref{sec3}, we then show that our particle-based BD model converges in a suitable limit to the PNP system~(\ref{NernstPlanck})--(\ref{Poisson}). 

In Section~\ref{sec4}, we present the details of computational implementations of both microscopic and macroscopic models. We discuss the strengths and limitations of each approach, applying them to an illustrative model comprising a system of Na$^+$ and Cl$^-$ ions. In Section~\ref{sec5}, we combine both approaches into a multi-resolution simulation framework, where BD simulations of individual ions and the mean-field PNP system~(\ref{NernstPlanck})--(\ref{Poisson}) are used in different parts of the computational domain. Such multi-resolution simulation frameworks have been developed in various contexts in the literature to simulate larger biological systems by using coarser (macroscopic) simulation approaches in parts of the computational domain~\cite{FleggChapman,Flegg:2015:CMC,Huber:2016:HFE,Robinson:2014:ATM,Franz:2013:MRA}. We conclude with the discussion of the applicability of the multi-resolution simulation techniques in Section~\ref{sec6}.

\section{Brownian dynamics} \label{sec2}
We model a system of $N$ positive and $N$ negative ions using BD in domain $\Omega \subset {\mathbb R}^{3}$. In our computational studies in Sections~\ref{sec4} and~\ref{sec5}, the domain $\Omega$ is cuboid 
\begin{equation}
\Omega = [0,L_1] \times [0,L_2] \times [0,L_3] \, ,
\label{cuboidomega}
\end{equation}
where $L_1$, $L_2$ and $L_3$ are positive parameters with units of length. The state of the system at time $t$ is described by the $6N$-dimensional vector
\begin{equation}
{\overline{\mathbf{X}}}(t)
=
\big[\mathbf{X}_1^+(t), \mathbf{X}_2^+(t), \dots, \mathbf{X}_N^+(t),\mathbf{X}_1^-(t), \mathbf{X}_2^-(t),\dots, \mathbf{X}_N^-(t)\big]
\in \Omega^{2N},
\label{statevector}
\end{equation}
where $\mathbf{X}_j^\pm(t)$ is given by~(\ref{ionpositions}) and denotes the position of the $j$-th positive ion~($\mathbf{X}_j^+$) or the $j$-th negative ion~($\mathbf{X}_j^-$) at time $t$ for $j=1,2,\dots,N$. The time evolution of  $\mathbf{X}_j^\pm(t)$ is given by the following It\^{o} stochastic differential equation
\begin{equation}
\mbox{d} \mathbf{X}_j^\pm(t)
= 
- 
\,
\alpha^\pm
\,
\nabla_j^\pm
U\big({\overline{\mathbf{X}}}\big)
\,
\mbox{d}t
+ 
\sqrt{2D^{\pm}} \,
\mbox{d} \mathbf{W}_j,
\qquad \mbox{for} \quad
j=1,2,\dots,N,
\label{BDSDEform}
\end{equation}
where  $\alpha^+$, $\alpha^-$, $D^+$ and $D^-$ are positive parameters, $U: {\mathbb R}^{6N} \to {\mathbb R}$ is the interaction potential and 
\begin{equation*}
\nabla_j^+
U
=
\left[
\frac{\partial U}{\partial x^+_{j,1}}, \,
\frac{\partial U}{\partial x^+_{j,2}}, \,
\frac{\partial U}{\partial x^+_{j,3}}
\right],
\quad
\nabla_j^-
U
=
\left[
\frac{\partial U}{\partial x^-_{j,1}}, \,
\frac{\partial U}{\partial x^-_{j,2}}, \,
\frac{\partial U}{\partial x^-_{j,3}}
\right].
\end{equation*}
Our BD model~(\ref{BDSDEform}) is a generalization of equation~(\ref{eq0}), where the interaction potential~$U$ is a function of $6N$ variables, because it depends not only on the position of the $j$-th positive or negative ion, but also on the positions of other ions in the system.

Our BD model~(\ref{BDSDEform}) can be derived in the overdamped limit from the Langevin dynamics~\cite{Erban:2020:SMR,Schuss:2001:DPN}. In particular, the parameter $\alpha^\pm$ is the reciprocal of the corresponding friction coefficient, which has units of [mass][time]$^{-1}$. Since the interaction potential~$U$ has units of energy, we deduce that our BD equation~(\ref{BDSDEform}) is dimensionally correct. However, to simplify our derivations of mean-field PNP equations, we first introduce the functional form of the interaction potential in Section~\ref{sec21} with a minimal number of parameters. In particular, our BD equation~(\ref{BDSDEform}) can be considered non-dimensionalized. Our parameters are related to physical quantities in Section~\ref{sec22}.

\subsection{Functional form of the interaction potential} \label{sec21} We assume that ions interact with fixed background potential and with each other. Moreover, we assume that our computational domain~(\ref{cuboidomega}) is a small representative part of a larger domain that contain ions. Such a process can be modelled using suitable boundary conditions at the boundary~$\partial \Omega$ of the cuboid. If the interactions were short-ranged, then ions leaving the domain~$\Omega$ could be removed from the simulation, while additional ions could be introduced at the domain boundary~$\partial \Omega$ according to a specified distribution~\cite{Erban:2014:MDB,Erban:2023:MDH}. However, interactions between ions include long-range (Coulomb) forces and the simulated ions (which are inside~$\Omega$) will also interact with ions which are outside of $\Omega$ and therefore not explicitly included in our simulation. To model such ions, we will implement standard periodic boundary conditions and assume that every ion interacts not only with the ions in $\Omega$, but also with their periodic image copies and with the periodic image copies of the background potential. To simplify our presentation, we introduce the notation
\begin{equation}
{\mathbf{L}}_{\boldsymbol{\ell}}
=
[\ell_1 \, L_1,\ell_2 \, L_2,\ell_3 \, L_3],
\qquad
\mbox{for any integer valued vector}
\quad
\boldsymbol{\ell}
=
[\ell_1,\ell_2,\ell_3]
\label{notation1}
\end{equation}
and we also denote summations over such integer valued vectors as$\,$:
\begin{equation}
\sum_{{\boldsymbol{\ell}}}
=
\sum_{\ell_1=-\infty}^{\infty}
\sum_{\ell_2=-\infty}^{\infty}
\sum_{\ell_3=-\infty}^{\infty}
\label{notation2}
\end{equation}
Denoting a point in the $6N$-dimensional state space by
\begin{equation}
{\overline{\mathbf{x}}}
=
\big[\mathbf{x}_1^+, \mathbf{x}_2^+, \dots, \mathbf{x}_N^+,\mathbf{x}_1^-, \mathbf{x}_2^-,\dots, \mathbf{x}_N^-\big],
\label{statespacepoint}
\end{equation}
the interaction potential $U\big({\overline{\mathbf x}}\big)$ is given as the following sum of five terms
\begin{eqnarray}
U\big({\overline{\mathbf x}}\big)
&=&
\sum_{\boldsymbol{\ell}}
\sum_{i=1}^N 
\left(
U_0^+ \big( \mathbf{x}^+_i - {\mathbf{L}}_{\boldsymbol{\ell}}\big)
+
U_0^- \big( \mathbf{x}^-_i - {\mathbf{L}}_{\boldsymbol{\ell}}\big)
+
\sum_{j=1}^{i-1}
U_1 \big( \!\parallel\! \mathbf{x}^+_i - \mathbf{x}^+_j - {\mathbf{L}}_{\boldsymbol{\ell}} \!\parallel\! \big)
\right.
\nonumber 
\\
&& +
\left.
\sum_{j=1}^{N}
U_2 \big( \!\parallel\! \mathbf{x}^-_i - \mathbf{x}^+_j - {\mathbf{L}}_{\boldsymbol{\ell}} \!\parallel\! \big)
+
\sum_{j=1}^{i-1}
U_3 \big( \!\parallel\! \mathbf{x}^-_i - \mathbf{x}^-_j - {\mathbf{L}}_{\boldsymbol{\ell}} \!\parallel\! \big)
\right),
\label{pairsum}
\end{eqnarray}
where $U_0^\pm : {\mathbb R}^3 \to {\mathbb R}$ is a background potential which includes interactions with fixed charges in the environment, $\parallel \!\!\cdots\!\!\parallel$ denotes the standard Euclidean norm, potential $U_1 : [0,\infty) \to {\mathbb R}$ (resp. $U_3 : [0,\infty) \to {\mathbb R}$) describes distance-dependent interactions between two positive ions (resp. two negative ions) and $U_2 : [0,\infty) \to {\mathbb R}$ describes interactions between a positive and a negative ion. The distance-dependent potentials $U_1,$ $U_2$ and $U_3$ have the same functional form given as the sum of the Lennard-Jones and Coulomb potentials. We write them as
\begin{equation}
U_k(r) = 
\frac{A_k}{r^{12}}
- 
\frac{B_k}{r^6}
+
\frac{C_k}{r}\,,
\label{eqpotential}
\end{equation}
where $A_k,$ $B_k$ and $C_k$, for $k=1,2,3$, are parameters, which we relate to physical quantities in Section~\ref{sec22}. We note that the sum in the third term of the interaction potential~(\ref{pairsum}) does not include the case $i=j$, which is missing for 
$\boldsymbol{\ell}
=
[0,0,0]$,
because the $i$-th ion does not interact with itself, and for $\boldsymbol{\ell}
\ne
[0,0,0]$
because the corresponding terms of the potential~(\ref{pairsum}) would be constant and they would not change the drift term of our BD model~(\ref{BDSDEform}), if we explicitly included them in~(\ref{pairsum}). In our illustrative computational simulations in Section~\ref{sec4}, we will implement the potential~(\ref{pairsum}) using the Ewald summation, which is described in Section~\ref{sec42}. 

Considering $r$ to be the distance between interacting particles with separation vector ${\mathbf x}$, we have $r=|{\mathbf x}|$ and equation~(\ref{eqpotential}) implies that
$\nabla^2 U_k = {\mathcal H}_k$ where
\begin{eqnarray}
{\mathcal H}_k(r)
&=&
\frac{132 \, A_k}{r^{14}}
- 
\frac{30 \, B_k}{r^8}
-
\frac{2 C_k \, \delta(r)}{r^2}
=
\frac{132 \, A_k}{r^{14}}
- 
\frac{30 \, B_k}{r^8}
-
4 \pi \, C_k \, \,\delta^3({\mathbf{x}})\, 
,
\label{laplaceeqpot}
\end{eqnarray}
where $\delta(\cdot)$ is the Dirac delta function and $\delta^3({\mathbf x})
=\delta(x_1) \, \delta(x_2) \, \delta(x_3)$ is the three-dimensional Dirac delta function.

\begin{table}
\begin{center}
\begin{tabular}{|l|l|l|}
\hline
parameter & value & reference \\
\hline\hline
\rule{0pt}{3.6mm} $A_1 \;$ &  
\rule{0pt}{4mm} \, $1.194 \times 10^{-8} \, \mbox{nm}^{12} \, \mbox{kcal} \, \mbox{mol}^{-1} \,$ 
& \cite{Hwang:2024:C4E} \\
\hline
\rule{0pt}{3.6mm} $B_1 \;$ & 
\rule{0pt}{4mm} \, $4.732 \times 10^{-5} \, \mbox{nm}^{6} \, \mbox{kcal} \, \mbox{mol}^{-1} \,$ & \cite{Hwang:2024:C4E} \\
\hline
\rule{0pt}{3.6mm} $A_2 \;$ & 
\rule{0pt}{4mm} \, $5.186 \times 10^{-7} \, \mbox{nm}^{12} \, \mbox{kcal} \, \mbox{mol}^{-1} \,$ & \cite{Hwang:2024:C4E} \\
\hline
\rule{0pt}{3.6mm} $B_2 \;$ & 
\rule{0pt}{4mm} \, 
$4.171 \times 10^{-4} \, \mbox{nm}^{6} \, \mbox{kcal} \, \mbox{mol}^{-1} \,$ & \cite{Hwang:2024:C4E} \\
\hline
\rule{0pt}{3.6mm} $A_3 \;$ &
\rule{0pt}{4mm} \, $1.150 \times 10^{-5} \, \mbox{nm}^{12} \, \mbox{kcal} \, \mbox{mol}^{-1} \,$  & \cite{Hwang:2024:C4E} \\
\hline
\rule{0pt}{3.6mm} $B_3 \;$ & 
\rule{0pt}{4mm} \, $2.627 \times 10^{-3} \, \mbox{nm}^{6} \, \mbox{kcal} \, \mbox{mol}^{-1} \,$ & \cite{Hwang:2024:C4E} \\
\hline
\rule{0pt}{3.6mm} $C_1 = - C_2 = C_3 \;$ & \rule{0pt}{4mm} \, $4.235 \times 10^{-1} \, \mbox{nm} \, \mbox{kcal} \, \mbox{mol}^{-1} \,$
&\cite{Fernandez:1997:FSP} and equation~(\ref{Clformula2}) \\
\hline
\rule{0pt}{3.6mm} $\alpha^+ \;$ & \rule{0pt}{4mm} \, $3.241 \times 10^{11} \, \mbox{kg}^{-1} \, \mbox{s}$ & \cite{Haynes:2014:CHC} and equation~(\ref{einstein}) \\
\hline
\rule{0pt}{3.6mm} $\alpha^- \;$ & \rule{0pt}{4mm} \, $
4.936 \times 10^{11} \, \mbox{kg}^{-1} \, \mbox{s}$  & \cite{Haynes:2014:CHC} and equation~(\ref{einstein}) \\
\hline
\rule{0pt}{3.6mm} $D^+ \;$ & \rule{0pt}{4mm} \, $1.334 \times 10^{9} \, \mbox{nm}^2 \, \mbox{s}^{-1}$ & \cite{Haynes:2014:CHC}  \\
\hline
\rule{0pt}{3.6mm} $D^- \;$ & \rule{0pt}{4mm} \, $2.032 \times 10^{9} \, \mbox{nm}^2 \, \mbox{s}^{-1}$ & \cite{Haynes:2014:CHC} \\
\hline
\end{tabular}
\end{center}
\vskip 2mm
\label{table1}
\caption{{\it \noindent Parameters of the potentials $U_1$, $U_2$ and $U_3$ given by~$(\ref{eqpotential})$, and parameters $\alpha^\pm$ and $D^\pm$ in the case of the positive ion being {\rm Na}$^{+}$ and the negative ion being {\rm Cl}$^{-}$.}}
\end{table}

\subsection{Parameterization} \label{sec22}
While our minimal set of parameters $A_k,$ $B_k$ and $C_k$, for $k=1,2,3,$ in~(\ref{eqpotential}) can be considered dimensionless for the purposes of the theoretical derivations presented in Section~\ref{sec3}, they can also be related to physical quantities and parameterized by the values in the literature.  In Sections~\ref{sec4} and~\ref{sec5}, we present illustrative simulations for the case when the positive ion is {\rm Na}$^{+}$ and the negative ion is {\rm Cl}$^{-}$. The parameters for these simulations are given in Table~\ref{table1}. The Lennard-Jones parameters $A_j$ and $B_j$ are calculated using the values in the CHARMM force field~\cite{Hwang:2024:C4E}. The parameters $\alpha^+$ and $\alpha^-$ are reciprocals of the corresponding friction coefficient, which can be estimated by different methods~\cite{Koneshan:1998:FCI}. We use the Einstein-Smoluchowski relation
\begin{equation}
\alpha^\pm = \frac{D^\pm}{k_b \, T}
\label{einstein}
\end{equation}
where $D^+$ (resp. $D^-$) is the diffusion constant of the positive (resp. negative) ion, $k_b$ is the Boltzmann constant and $T$ is the absolute temperature.
In Table~\ref{table1}, we use the values of $D^+$ and $D^-$ at temperature $T=25\,^{\circ}{\hskip -0.3mm}\mathrm{C}$ for sodium and calcium ions~\cite{Haynes:2014:CHC}, repectively, and we calculate $\alpha^\pm$ by~(\ref{einstein}).
Since $C_k$ corresponds to the Coulomb term, we have 
\begin{equation}
C_1 = \frac{(q^+)^2}{4 \pi \varepsilon_0 \, \varepsilon}\,,
\qquad
C_2 = \frac{q^+q^-}{4 \pi \varepsilon_0 \,\varepsilon}
\qquad
\mbox{and}
\qquad
C_3 = \frac{(q^-)^2}{4 \pi \varepsilon_0 \,\varepsilon}\,,
\label{Clformula}
\end{equation}
where $q^+$ (resp. $q^-$) is the charge of the positive (resp. negative) ion, $\varepsilon_0$ is the dielectric permittivity constant of the vacuum and $\varepsilon$ is the dielectric permittivity constant of the medium. Since we consider monovalent ions {\rm Na}$^{+}$ and {\rm Cl}$^{-}$, we have $q^+={\mathrm e}$ and $q^-=-{\mathrm e}$ in formulas~(\ref{Clformula}), where
${\mathrm e}=1.602 \times 10^{-19} \, {\mathrm C}$ is the elementary charge. Consequently, formulas~(\ref{Clformula}) imply
\begin{equation}
C_1 = - C_2 = C_3 = \frac{{\mathrm e}^2}{4 \pi \varepsilon_0 \, \varepsilon}\,,
\label{Clformula2}
\end{equation}
which we use in our illustrative simulations together with $\varepsilon$ being the permittivity of water at temperature $T=25\,^{\circ}{\hskip -0.3mm}\mathrm{C}$ given as~$\varepsilon=78.41$ in the literature~\cite{Fernandez:1997:FSP}. 

\section{Derivation of the PNP system from the BD model} \label{sec3}

In this section, we start with the BD model~(\ref{BDSDEform}) and we derive
the PNP system~(\ref{NernstPlanck})--(\ref{Poisson}) in a suitable macroscopic limit. The derived system is used in Section~\ref{sec5} to design a  multi-resolution computational method. The PNP system has been previously derived in the literature starting from a microscopic model written in terms of the Langevin dynamics~\cite{Schuss:2001:DPN}. Alternatively, the PNP system can also be derived from the alterations in the free energy functional, which encompasses both the electrostatic free energy and the ideal component of the chemical potential~\cite{Gillespie:2002:CPN,Lu:2010:PNP}.

\subsection{Derivation of the Poisson equation}

The interaction potential~$U$ in our formulation~(\ref{BDSDEform}) of the BD model is a function of $6N$ variables given by~(\ref{pairsum}), while the potential $\phi$ in the Poisson equation~(\ref{Poisson}) is a function of three spatial coordinates. To derive the PNP system, we first rewrite the BD model~(\ref{BDSDEform}) using potentials $\Phi^\pm: {\mathbb R}^3 \to {\mathbb R}$ that depend on three spatial variables and are parametrized by current positions of ions ${\overline{\mathbf{X}}}(t)$ given by~(\ref{statevector}). We define
\begin{eqnarray*}
\Phi^+\big({\mathbf x}\big)
&=&
\sum_{\boldsymbol{\ell}}
\left(
U_0^+ \big( \mathbf{x} - {\mathbf{L}}_{\boldsymbol{\ell}}\big)
+
\sum_{i=1}^{N}
\Big(
U_1 \big( \!\parallel\! \mathbf{x} -\! \mathbf{X}^+_i \!- {\mathbf{L}}_{\boldsymbol{\ell}} \!\parallel\! \big)
+
U_2 \big( \!\parallel\! \mathbf{x} -\! \mathbf{X}^-_i \!- {\mathbf{L}}_{\boldsymbol{\ell}} \!\parallel\! \big)
\!\Big)\!\!
\right)\!,
\\
\Phi^-\big({\mathbf x}\big)
&=&
\sum_{\boldsymbol{\ell}}
\left(
U_0^- \big( \mathbf{x} - {\mathbf{L}}_{\boldsymbol{\ell}}\big)
+
\sum_{i=1}^{N}
\Big(
U_2 \big( \!\parallel\! \mathbf{x} -\! \mathbf{X}^+_i \!- {\mathbf{L}}_{\boldsymbol{\ell}} \!\parallel\! \big)
+
U_3 \big( \!\parallel\! \mathbf{x} -\! \mathbf{X}^-_i \!- {\mathbf{L}}_{\boldsymbol{\ell}} \!\parallel\! \big)\!\Big)\!\!\right)\!.
\end{eqnarray*}
We note that the potential $\Phi^\pm({\mathbf x})$ is singular for ${\mathbf x} = \mathbf{X}^\pm_j$, $j=1,2,\dots,N$, so we also define potentials $\Phi^\pm_j: {\mathbb R}^3 \to {\mathbb R}$ that remove this singularity and all terms depending on $\mathbf{X}^\pm_j$ by
\begin{eqnarray}
\Phi^+_j\big({\mathbf x}\big)
&=&
\Phi^+\big({\mathbf x}\big)
-
\sum_{\boldsymbol{\ell}}
U_1 \big( \!\parallel\! \mathbf{x} - \mathbf{X}^+_j \!- {\mathbf{L}}_{\boldsymbol{\ell}}  \!\parallel\! \big)\,,
\label{nonsingpotentialsplus}    
\\
\Phi^-_j\big({\mathbf x}\big)
&=&
\Phi^-\big({\mathbf x}\big)
-
\sum_{\boldsymbol{\ell}}
U_3 \big( \!\parallel\! \mathbf{x} - \mathbf{X}^-_j  \!- {\mathbf{L}}_{\boldsymbol{\ell}} \!\parallel\! \big)\,.
\label{nonsingpotentialsminus}    
\end{eqnarray}
Then our BD model~(\ref{BDSDEform}) can be rewritten as
\begin{equation}
\mbox{d} \mathbf{X}_j^\pm(t)
= 
- \alpha^\pm
\,
\nabla
\left(
\Phi^\pm_j \big({\mathbf{X}}_j^\pm\big)
\right)
\mbox{d}t
+ 
\sqrt{2D^{\pm}} \,
\mbox{d} \mathbf{W}_j\,,
\label{BDSDEform3D}
\end{equation}
for $j=1,2,\dots,N.$ Using our notation~(\ref{laplaceeqpot}), we observe that the potential $\Phi^\pm({\mathbf x})$ satisfies the Poisson equation in the form
\begin{eqnarray*}
\nabla^2 \Phi^+\big({\mathbf x}\big)
&=&
\sum_{\boldsymbol{\ell}}
{\mathcal H}_0^+
\big(\mathbf{x}- {\mathbf{L}}_{\boldsymbol{\ell}}\big)
+
\sum_{i=1}^{N}
\Big(
{\mathcal H}_1 \big( \!\parallel\! \mathbf{x} -\! \mathbf{X}^+_i \!- {\mathbf{L}}_{\boldsymbol{\ell}} \!\parallel\! \big)
+
{\mathcal H}_2 \big( \!\parallel\! \mathbf{x} - \!\mathbf{X}^-_i \!- {\mathbf{L}}_{\boldsymbol{\ell}} \!\parallel\! \big)
\Big),
\\
\nabla^2 \Phi^-\big({\mathbf x}\big)
&=&
\sum_{\boldsymbol{\ell}}
{\mathcal H}_0^-
\big(\mathbf{x}- {\mathbf{L}}_{\boldsymbol{\ell}}\big)
+
\sum_{i=1}^{N}
\Big(
{\mathcal H}_2 \big( \!\parallel\! \mathbf{x} - \!\mathbf{X}^+_i \!- {\mathbf{L}}_{\boldsymbol{\ell}} \!\parallel\! \big)
+
{\mathcal H}_3 \big( \!\parallel\! \mathbf{x} - \!\mathbf{X}^-_i\! - {\mathbf{L}}_{\boldsymbol{\ell}} \!\parallel\! \big)\Big),
\end{eqnarray*}
where we have defined ${\mathcal H}_0^\pm = \nabla^2 U_0^\pm.$
In particular, we have formally obtained the Poisson equation, which includes the current positions of all ions. To get equation~(\ref{Poisson}) for the electrostatic potential, we focus on the terms on the right-hand side, which correspond to the electrostatic (Coulomb) forces. The Coulomb terms are included in equation~(\ref{laplaceeqpot}) as terms corresponding to constant $C_k$ We also assume that there is an underlying permanent charge density $\varrho_{\mathrm{p}}$ which helps us to express ${\mathcal H}_0^\pm$ as
\begin{equation}
{\mathcal H}_0^\pm (\mathbf{x})
= 
\frac{q^\pm \varrho_{\mathrm{p}}(\mathbf{x})}{\varepsilon_0 \, \varepsilon}
+
{\mathcal H}_{0,\mathrm{LJ}}^\pm(\mathbf{x})\,,
\label{defvarrhoperm}    
\end{equation}
where the extra term ${\mathcal H}_{0,\mathrm{LJ}}^\pm(\mathbf{x})$ corresponds to possible short-range interactions, which include some Lennard-Jones terms in our illustrative simulations in Section~\ref{sec4}. The term ${\mathcal H}_{0,\mathrm{LJ}}^\pm(\mathbf{x})$ does not appear in the Poisson equation~(\ref{potentialxovery}) below, but it is included in the error term~(\ref{errordefpmj}) in Section~\ref{sec32}.

Using our notation~(\ref{statespacepoint}) for points in the $6N$-dimensional state space, we define the electrostatic potential as three-dimensional vector field $\widehat{\phi}\big(\,\cdot\,; \overline{\mathbf{y}}\big) : {\mathbb R}^3 \to {\mathbb R}$ parameterized by the $6N$-dimensional state vector $\overline{\mathbf{y}} \in \Omega^{2N}$. It is given as the solution of the Poisson equation
\begin{equation}
\; \nabla^2 \widehat{\phi}\big(\mathbf{x}; \overline{\mathbf{y}}\big) 
= -\frac{1}{\varepsilon_0 \, \varepsilon} \sum_{\boldsymbol{\ell}}
\left[
\varrho_{\mathrm{p}}(\mathbf{x}\!- \!{\mathbf{L}}_{\boldsymbol{\ell}}) 
+ 
\sum_{i=1}^{N}
\Big(
q^+ \,
\delta^3
\big(\mathbf{x}\!-\!\mathbf{y}_i^+
\!\!-\! {\mathbf{L}}_{\boldsymbol{\ell}}\big) 
+ 
q^- \,
\delta^3
\big(\mathbf{x}\!-\!\mathbf{y}_i^-
\!\!-\! {\mathbf{L}}_{\boldsymbol{\ell}}\big)\!
\Big)\!
\right]\!,
\label{potentialxovery}
\end{equation}
which has the form of the Poisson equation~(\ref{Poisson}) with concentrations replaced by the sums of the Dirac delta functions. Substituting the state space position $\overline{\mathbf{X}}(t)$ of ions at time $t$ for $\overline{\mathbf{y}}$, we obtain that~$\widehat{\phi}\big(\,\cdot\,; \overline{\mathbf{X}}(t) \big) : {\mathbb R}^3 \to {\mathbb R}$ is a three-dimensional vector field which captures all Coulomb terms in our BD potential~(\ref{nonsingpotentialsplus})--(\ref{nonsingpotentialsminus}).

Denoting a point in the $6N$-dimensional state space by~(\ref{statespacepoint}) and the $6N$-dimensio\-nal volume differential by
\begin{equation}
\mbox{d} {\overline{\mathbf{x}}}
=
\mbox{d} \mathbf{x}_1^+ \, \mbox{d} \mathbf{x}_2^+ \, \dots \, \mbox{d} \mathbf{x}_N^+ \, \mbox{d} \mathbf{x}_1^- \, \mbox{d} \mathbf{x}_2^- \, \dots \, \mbox{d} \mathbf{x}_N^- \,,
\label{intpoint}
\end{equation}
we can equivalently describe our BD model~(\ref{BDSDEform3D}) by the time dependent probability density function $p: \Omega^{2N} \times [0,\infty) \to [0,\infty)$ which is defined so that 
$$
p\big({\overline{\mathbf{x}}},t\big) \, \mbox{d} {\overline{\mathbf{x}}} \quad \mbox{is the probability that} \quad {\overline{\mathbf{X}}}(t) \in [\overline{\mathbf{x}},\overline{\mathbf{x}}+\mbox{d}\overline{\mathbf{x}}).
$$
Using our notation~(\ref{statespacepoint}) for points in the $6N$-dimensional state space, we further simplify our presentation by denoting
\begin{eqnarray}
\widetilde{\mathbf{x}}^+_j
 & = & 
(\mathbf{x}_1^+, \mathbf{x}_2^+, \dots, \mathbf{x}_{j-1}^+, \mathbf{x}_{j+1}^+, \dots,\mathbf{x}_N^+,\mathbf{x}_1^-, \mathbf{x}_2^-,\dots, \mathbf{x}_N^-)\,,
\label{plustilde}
\\     
\widetilde{\mathbf{x}}^-_j
 & = &  (\mathbf{x}_1^+, \mathbf{x}_2^+, \dots, \mathbf{x}_N^+, \mathbf{x}_1^-, \mathbf{x}_2^-, \dots, \mathbf{x}_{j-1}^-, \mathbf{x}_{j+1}^-,\dots, \mathbf{x}_N^-)\,,
 \label{negtilde}
\end{eqnarray}
that is, $\widetilde{\mathbf{x}}^+_j \in \Omega^{2N-1}$ $\big($resp. $\widetilde{\mathbf{x}}^-_j \in \Omega^{2N-1}\big)$ is a projection of the $6N$-dimensional state space on the $(6N-3)$-dimensional subspace which excludes the position of the $j$-th positive (resp. negative) ion, for $j=1,2,\dots,N$. Using notation~(\ref{plustilde})--(\ref{negtilde}), the concentration $c^\pm(\mathbf{x},t)$ is given as the sum of marginal densities of each ion being at the position $\mathbf{x}$ at time $t$. Since ions are indistinguishable, we can write the concentration $c^\pm(\mathbf{x},t)$ as $N$ times the marginal density of the first ion in the list, i.e.
\begin{equation}
c^\pm\big(\mathbf{x},t\big) = N 
\int_{ \Omega^{2N-1}} p \Big(\mathbf{x},\widetilde{\mathbf{x}}^\pm_1,t\Big) \, \mbox{d}
\widetilde{\mathbf{x}}^\pm_1.
\label{defcmarg}
\end{equation}
Multiplying equation~(\ref{potentialxovery}) by $p\big({\overline{\mathbf{y}}},t\big)$ and integrating over $\overline{\mathbf{y}} \in \Omega^{2N}$, we obtain
\begin{equation}
\; \nabla^2 \phi\big(\mathbf{x}\big) 
= -\frac{1}{\varepsilon_0 \,\varepsilon} \sum_{\boldsymbol{\ell}}
\Big[
\varrho_{\mathrm{p}}(\mathbf{x}- {\mathbf{L}}_{\boldsymbol{\ell}}) 
\,+\, 
q^+ \,
c^+
\big(\mathbf{x} - {\mathbf{L}}_{\boldsymbol{\ell}}\big) 
\,+\, 
q^- \,
c^-
\big(\mathbf{x} - {\mathbf{L}}_{\boldsymbol{\ell}}\big)
\Big]\!,
\label{potentialxintegratedy}
\end{equation}
where $\phi\big(\mathbf{x}\big)$ is an averaged electrostatic potential give by
\begin{equation}
\phi\big(\mathbf{x}\big)
=
\int_{\Omega^{2N}}
\widehat{\phi}\big(\mathbf{x}; \overline{\mathbf{y}}\big) 
\, p\big({\overline{\mathbf{y}}},t\big) \, \mbox{d} \overline{\mathbf{y}}.
\label{averelpotnial}
\end{equation}
Since the summation~(\ref{notation2}) over integer valued vectors~(\ref{notation1}) periodically covers ${\mathbb R}^3$, we can equivalently rewrite equation~(\ref{potentialxintegratedy}) as equation
\begin{equation}
\; \nabla^2 \phi\big(\mathbf{x}\big) 
= -\frac{1}{\varepsilon_0 \,\varepsilon} \Big[
\varrho_{\mathrm{p}}(\mathbf{x}) 
\,+\, 
q^+ \,
c^+
\big(\mathbf{x}\big) 
\,+\, 
q^- \,
c^-
\big(\mathbf{x}\big)
\Big]\!,
\label{potentialxintegratedyper}
\end{equation}
which is solved in cuboid domain~(\ref{cuboidomega}) with periodic boundary conditions. In particular, we observe that the averaged electrostatic potential~(\ref{averelpotnial}) satisfies the Poisson equation~(\ref{Poisson}) in domain $\Omega.$

\subsection{Derivation of the Nernst-Planck equations} 
\label{sec32}

At time $t$, the electrostatic potential corresponding to both ions and permanent charge density is given as~$\widehat{\phi}\big(\mathbf{x}; \overline{\mathbf{X}}(t)\big)$, where $\overline{\mathbf{X}}(t)$ is the state of the system at time $t$ and~$\widehat{\phi}\big(\mathbf{x}; \overline{\mathbf{y}}\big)$ is the solution to the Poisson equation~(\ref{potentialxovery}). When using this potential in the evolution equation~(\ref{BDSDEform3D}) for the $j$-th positive or negative ion, we need to remove the potential corresponding to this ion. Therefore, we also define potentials
\begin{equation}
\widehat{\phi}_j^\pm \big(\mathbf{x}; \widetilde{\mathbf{y}}^\pm_j\big)
=
\widehat{\phi}\big(\mathbf{x}; \overline{\mathbf{y}}\big)
-
\sum_{\boldsymbol{\ell}}
\frac{q^\pm}{4 \pi \varepsilon_0 \, \varepsilon} \frac{1}{\!\parallel\! \mathbf{x} \!- \!\mathbf{y}^\pm_j \!-\! {\mathbf{L}}_{\boldsymbol{\ell}}\!\parallel\! } \,,
\label{reducedpotential}
\end{equation}
where we use notation~(\ref{plustilde})--(\ref{negtilde}) to indicate that the left hand side is parameterized by the $(6N\!-\!3)$--dimensional vector~$\widetilde{\mathbf{y}}^\pm_j$. The potential $\widehat{\phi}_j^\pm\big(\mathbf{x}; \widetilde{\mathbf{y}}^\pm_j\big)$ does not capture all terms of the interaction potential $U\big({\overline{\mathbf x}}\big)$ given in~(\ref{pairsum}). To rewrite the BD model~(\ref{BDSDEform3D}) using the averaged electrostatic potential~$\phi(\textbf{x})$, 
we define the error term as the difference
\begin{equation}
E^\pm_j\big({\mathbf x};  \widetilde{\mathbf{X}}_j^\pm \big)
=
\Phi^\pm_j\big({\mathbf x}\big)
\,-\,
q^\pm \, \widehat{\phi}^\pm_j\big(\textbf{x};  \widetilde{\mathbf{X}}_j^\pm\big)\,,
\label{errordefpmj}    
\end{equation}
where $\Phi^\pm_j\big({\mathbf x}\big)
\equiv \Phi^\pm_j\big({\mathbf x};  \widetilde{\mathbf{X}}_j^\pm \big)$ is given by (\ref{nonsingpotentialsplus})--(\ref{nonsingpotentialsminus}). Substituting~(\ref{errordefpmj}) into equation~(\ref{BDSDEform3D}), we reformulate our BD~model as
\begin{equation}
\mbox{d} \mathbf{X}_j^\pm(t)
= 
- 
\,
\alpha^\pm
\,
\nabla
\left(
q^\pm 
\,\widehat{\phi}_j^\pm\big({{\textbf X}_j^\pm};  \widetilde{\mathbf{X}}_j^\pm\big)
+
E^\pm_j\big({{\mathbf X}_j^\pm};  \widetilde{\mathbf{X}}_j^\pm
\big)
\right)
\,
\mbox{d}t
+ 
\sqrt{2D^{\pm}} \,
\mbox{d} \mathbf{W}_j\,,
\label{BDSDEform3Delpot}
\end{equation}
for $j=1,2,\dots,N.$ The time evolution of $p\big({\overline{\mathbf{x}}},t\big)$ is given by the Fokker-Planck equation~\cite{vanKampen:2007:SPP,Mao:2007:SDE}
\begin{eqnarray}
\frac{\partial p}{\partial t}\big({\overline{\mathbf{x}}},t\big) 
&=&
\sum_{j=1}^N
\nabla_j^+
\cdot
\left(
D^+ \nabla_j^+ p
\,+\,
p
\,
\alpha^+
\nabla_j^+
\left(
q^+ 
\,\widehat{\phi}_j^+\big({{\textbf x}_j^+};  \widetilde{\mathbf{x}}_j^+\big)
+
E^+_j\big({{\mathbf x}_j^+};  \widetilde{\mathbf{x}}_j^+
\right)
\right)
\nonumber
\\ && +
\sum_{j=1}^N
\nabla_j^-
\cdot
\left(
D^- \nabla_j^- p
\,+\,
p
\,
\alpha^-
\nabla_j^-
\left(
q^-
\,\widehat{\phi}_j^-\big({{\textbf x}_j^-};  \widetilde{\mathbf{x}}_j^-\big)
+
E^-_j\big({{\mathbf x}_j^-};  \widetilde{\mathbf{x}}_j^-
\right)
\right).
\label{FokPlan6N}
\end{eqnarray}
Multiplying by $N$, integrating over~$\widetilde{\mathbf{x}}^\pm_1$, and using~(\ref{defcmarg}), (\ref{averelpotnial}) and~(\ref{einstein}), we obtain
\begin{equation}
\frac{\partial c^\pm}{\partial t} 
=
\nabla
\cdot
D^\pm 
\left(
\nabla c^\pm
\,+\,
\frac{q^\pm}{k_b T}
\,
c^\pm
\,
\nabla \phi
\,+\,
\mathbf{f}_{\mathrm{er}}
\right),
\label{fokkplanck}
\end{equation}
where the error term is given by
\begin{eqnarray}
\mathbf{f}_{\mathrm{er}} \big(\mathbf{x},t\big) 
&=&  \frac{N}{k_b T} 
\int_{ \Omega^{2N-1}} p \big(\mathbf{x},\widetilde{\mathbf{x}}^\pm_1,t\big) \,\nabla_1^\pm E_1^\pm \big({{\textbf x}};  \widetilde{\mathbf{x}}_1^\pm \big) \,  
    \mbox{d} \widetilde{\mathbf{x}}^\pm_1    \nonumber\\
&+&  \frac{N q^\pm}{k_b T} 
\int_{ \Omega^{2N-1}} \int_{\Omega^{2N}} p \big(\mathbf{x},\widetilde{\mathbf{x}}^\pm_1,t\big) \, p\big({\overline{\mathbf{y}}},t\big) \,
\left(
\nabla_1^\pm \widehat{\phi}_1^\pm \big({{\textbf x}};  \widetilde{\mathbf{x}}_1^\pm \big) 
-
\nabla \widehat{\phi}\big(\mathbf{x}; \overline{\mathbf{y}}\big) 
\right)\, \mbox{d} \overline{\mathbf{y}}\,  
    \mbox{d} \widetilde{\mathbf{x}}^\pm_1. 
  \label{errorforce}
\end{eqnarray}
Consequently, equation~(\ref{fokkplanck}) reduces to the Nernst-Planck equations~(\ref{NernstPlanck}) provided that 
$\mathbf{f}_{\mathrm{er}} \equiv \mathbf{0}.$

\section{Comparison of BD with the solution obtained by the PNP system} \label{sec4}

In this section, we will implement both the microscopic and macroscopic descriptions of an illustrative model comprising a system of Na$^+$ and~Cl$^-$ ions. The macroscopic description is given by the PNP equations~(\ref{NernstPlanck})--(\ref{Poisson}) derived in Section~\ref{sec3}. This is followed in Section~\ref{sec5} by developing a multi-resolution scheme that uses both BD and PNP in the same dynamical simulation.

\begin{figure}
\rule{0pt}{1pt}
\raise 2mm
\hbox{\epsfig{file=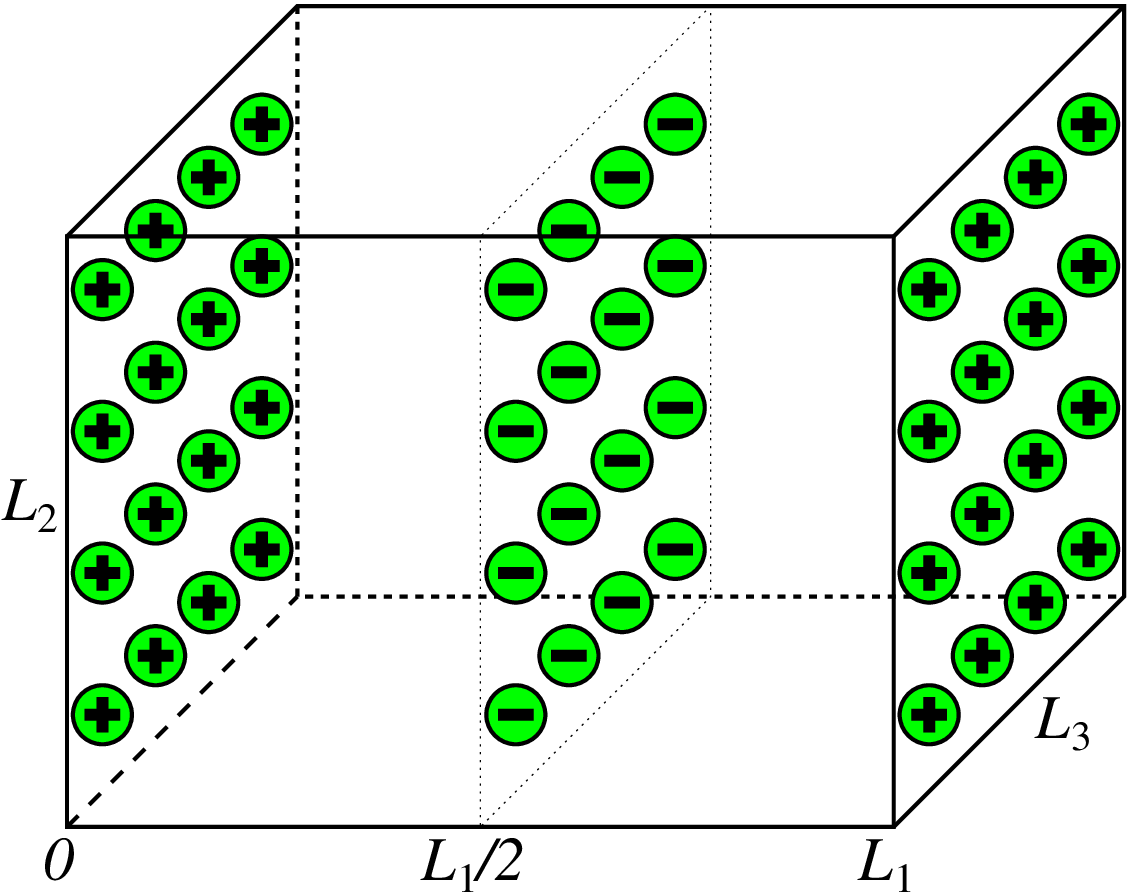,height=4.45cm}}
\hskip 4mm \epsfig{file=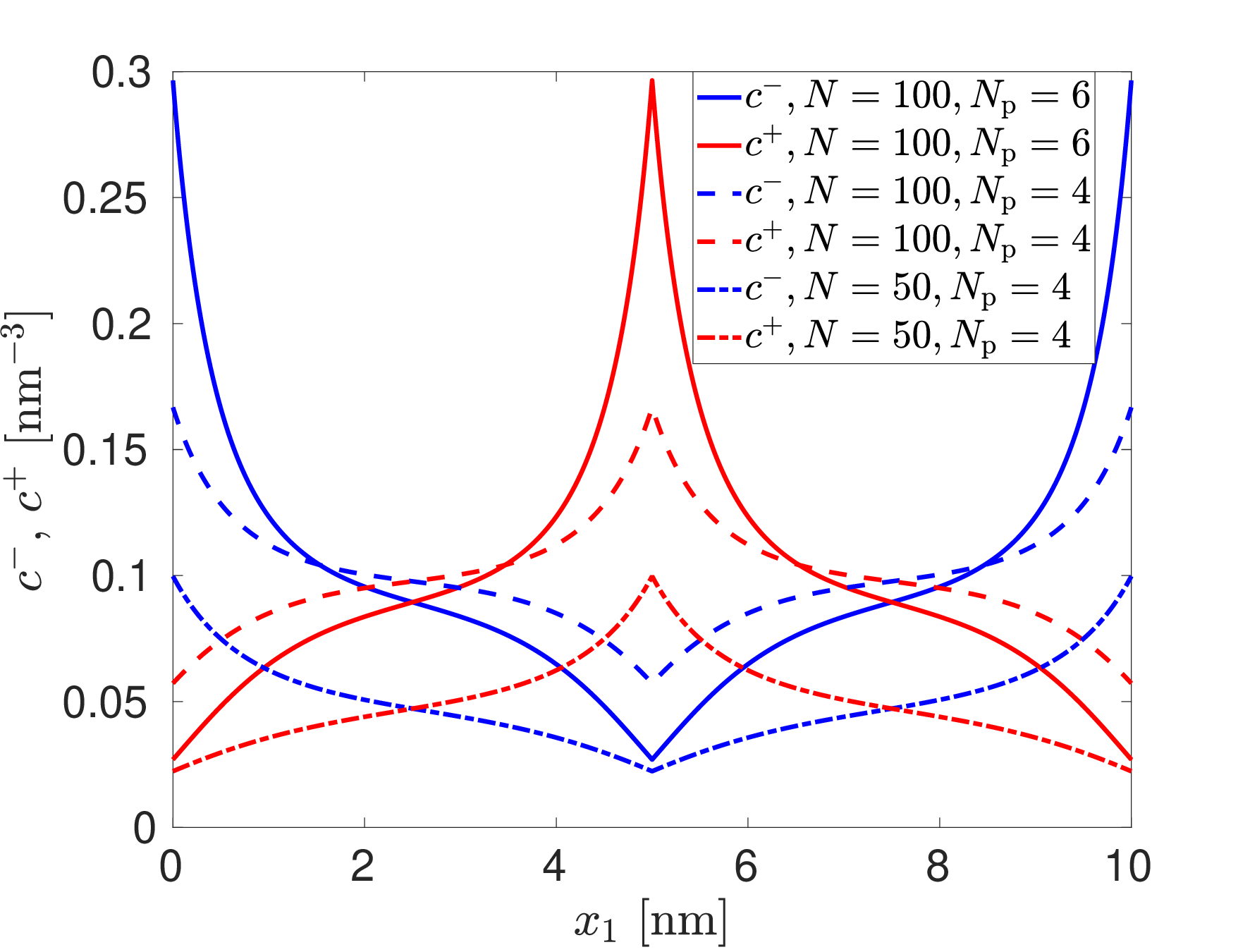,height=5.2cm}
\vskip -5.4cm
(a) \hskip 5.4cm (b)
\vskip 4.7cm
\caption{{\rm (a)} Schematic of the computational domain $\Omega$ given by~$(\ref{cuboidomega})$ with periodic lattices of point charges on planes $x_1 = 0$, $x_1 = L_1/2$ and $x_1=L_1$. We use $N_{\mathrm{p}}=4.$ \hfill\break {\rm (b)} The steady state solutions $c^-(x_1)$ and $c^+(x_1)$ obtained by one-dimensional approximation~$(\ref{Boltzmann})$ for three different choices of $N$ and $N_{\mathrm{p}}$, namely: (i) $N=100$ and $N_{\mathrm{p}}=6$ (solid lines); \hfill\break (ii) $N=100$ and $N_{\mathrm{p}}=4$ (dashed lines); and (iii) $N=50$ and $N_{\mathrm{p}}=4$ (dot-dashed lines).}
\label{fig1}
\end{figure}

We consider a simulation of $N$ sodium and $N$ chloride ions in domain $\Omega$ defined by~(\ref{cuboidomega}). In particular, we use the parameters given for~Na$^+$ and~Cl$^-$ in Table~\ref{table1}. The permanent charge density $\varrho_{\mathrm{p}}$, which appears in~(\ref{Poisson}) or~(\ref{defvarrhoperm}), is given in our example as regular rectangular lattices of permanent positive (on plane $x_1=0$) and negative charges (on plane $x_1 = L_1/2$). Since we consider periodic boundary conditions, these layers are repeated in all planes $x_1=j \, L_1$ and  $x_1=(j+1/2) \, L_1$, where $j$ in an integer. Our domain is schematically shown in Figure~\ref{fig1}(a), where we highlight the layer of permanent charges in planes $x_1 = 0$, $x_1 = L_1/2$ and $x_1=L_1$. More precisely, given an integer value $N_{\mathrm{p}}$, we have $N_{\mathrm{p}}^2$ elementary charges on the side $x_1=0$ of the cuboid~(\ref{cuboidomega}) at points
\begin{equation}
\mathbf{z}_{m,n}^+
=
\left[
0, \, \frac{(m-1/2)L_2}{N_{\mathrm{p}}}, \, \frac{(n-1/2)L_3}{N_{\mathrm{p}}} 
\right]
\qquad
\mbox{for} \quad m,n=1, 2, \dots, N_{\mathrm{p}}\,,
\label{permNachargeslocations}
\end{equation}%
and we also have $N_{\mathrm{p}}^2$ negative elementary charges on the side $x_1=L_1/2$ of the cuboid~(\ref{cuboidomega}) at points 
\begin{equation}
\mathbf{z}_{m,n}^-
=
[L_1/2,0,0]
+
\mathbf{z}_{m,n}^+
\qquad \mbox{for} \quad
m,n=1, 2, \dots, N_{\mathrm{p}}.
\label{permClchargeslocations}
\end{equation}
Therefore the permanent charge density is given as
\begin{equation}
\varrho_{\mathrm{p}}\big(\mathbf{x}\big)
=
\sum_{m=1}^{N_{\mathrm{p}}}
\sum_{n=1}^{N_{\mathrm{p}}}
{\mathrm{e}} \, 
\delta^3 \!
\left(
\mathbf{x} - 
\mathbf{z}_{m,n}^+
\right)
-
{\mathrm{e}} \, 
\delta^3 \!
\left(
\mathbf{x} - 
\mathbf{z}_{m,n}^-
\right).
\label{varrhopexample}
\end{equation}
In our model, we identify the elementary charges on the side $x_1=0$ with Na$^+$ ions at fixed positions~(\ref{permNachargeslocations}) and the elementary charges on the side $x_1=L/2$ with Cl$^-$ ions at fixed positions~(\ref{permClchargeslocations}). In particular, ions are repelled from the fixed charges at short distances by the corresponding Lennard-Jones type interactions, giving the Laplacian of the background potential~(\ref{defvarrhoperm}) in the form 
\begin{eqnarray*}
{\mathcal H}_0^+ (\mathbf{x})
&=& 
\frac{q^+ \varrho_{\mathrm{p}}(\mathbf{x})}{\varepsilon_0 \, \varepsilon}
\!+\!
\sum_{m,n=1}^{N_{\mathrm{p}}}
\frac{132 \, A_1}{\!\parallel\! \mathbf{x} \!-\! \mathbf{z}^+_{m,n} \!\!\parallel\!^{\!14}}
\!- 
\frac{30 \, B_1}{\!\parallel\! \mathbf{x} \!-\! \mathbf{z}^+_{m,n} \!\!\parallel\!^{\!8}}\,
\!+
\frac{132 \, A_2}{\!\parallel\! \mathbf{x} \!-\! \mathbf{z}^-_{m,n} \!\!\parallel\!^{\!14}}
\!- 
\frac{30 \, B_2}{\!\parallel\! \mathbf{x} \!-\! \mathbf{z}^-_{m,n} \!\!\parallel\!^{\!8}},
\\   
{\mathcal H}_0^- (\mathbf{x})
&=& 
\frac{q^- \varrho_{\mathrm{p}}(\mathbf{x})}{\varepsilon_0 \, \varepsilon}
\!+\!
\sum_{m,n=1}^{N_{\mathrm{p}}}
\frac{132 \, A_2}{\!\parallel\! \mathbf{x} \!-\! \mathbf{z}^+_{m,n} \!\!\parallel\!^{\!14}}
\!- 
\frac{30 \, B_2}{\!\parallel\! \mathbf{x} \!-\! \mathbf{z}^+_{m,n} \!\!\parallel\!^{\!8}}\,
\!+
\frac{132 \, A_3}{\!\parallel\! \mathbf{x} \!-\! \mathbf{z}^-_{m,n} \!\!\parallel\!^{\!14}}
\!-
\frac{30 \, B_3}{\!\parallel\! \mathbf{x} \!-\! \mathbf{z}^-_{m,n} \!\!\parallel\!^{\!8}},
\end{eqnarray*}
where parameters $A_1$ and $B_1$ (resp. $A_2$ and $B_2$,
or $A_3$ and $B_3$) correspond to Na$^+$-Na$^+$ (resp. Na$^+$-Cl$^-$ or Cl$^-$-Cl$^-$) interactions and are given in Table~\ref{table1}. The cuboid domain~(\ref{cuboidomega}) is chosen with equal parameters 
$L_1 = L_2 = L_3 = 10 \, \mbox{nm},$
i.e. $\Omega$ is a cube and we implement periodic boundary conditions as discussed in Section~\ref{sec2}. Our model will be investigated using BD~simulations in Section~\ref{sec42} and by numerically solving the PNP system~(\ref{NernstPlanck})--(\ref{Poisson}) in Section~\ref{sec43}. We begin by discussing the equilibrium properties of our model, when the PNP system reduces to solving the Poisson-Boltzmann equation. 

\subsection{One-dimensional approximation} 

\label{sec41}

In Figure~\ref{fig1}(a), permanent charges are placed on a rectangular lattice with spacing $L_2/N_p$ and $L_3/N_p$. To get some insight into the model behaviour, we first approximate the layers of sodium and chloride ions at positions~(\ref{permNachargeslocations}) and~(\ref{permClchargeslocations}) as uniformly charged planes at $x_1=0$ and $x_1=L_1/2$ with surface charge density $\sigma^+$ and $\sigma^-$, respectively, where
$$
\sigma^\pm
=
\pm \frac{N_{\mathrm{p}}^2 \, \mathrm{e}}{L_2 \, L_3}.
$$
Using this assumption, we can develop a one-dimensional approximation of our system, because $c^\pm$ and $\phi$ are constant in $x_2$ and $x_3$ directions. Considering the equilibrium properties of our system, we can substitute
\begin{equation}
c^\pm(\mathbf{x}) = c^\pm(x_1)
\qquad
\mbox{and} 
\qquad \phi(\mathbf{x}) = \phi(x_1)
\label{x1dependence}
\end{equation}
into the Nernst-Planck equation~(\ref{NernstPlanck}) and pass $t \to \infty$ to get
\begin{equation}
c^\pm(x_1)
=
c_0 \, \exp \! \left(
-
\frac{q^\pm\, \phi(x_1)}{k_b T}
\right),
\label{Boltzmann}
\end{equation}
where $c_0$ is a constant given by the normalization condition
\begin{equation}
N
=
\int_0^{L_1} \!\!\!
\int_0^{L_2} \!\!\!
\int_0^{L_3} \!\!
c^\pm(\mathbf{x})
\, \mbox{d} x_3
\, \mbox{d} x_2
\, \mbox{d} x_1
=
L_2 \, L_3
\int_{0}^{L_1}
c^\pm(x_1) \, \mbox{d} x_1.
\label{normcondition}
\end{equation}
Substituting~(\ref{x1dependence}) and~(\ref{Boltzmann}) into the Poisson equation~(\ref{Poisson}), and using $q^+ = \mathrm{e}$ and $q^- = -\mathrm{e}$, we get
\begin{equation}
\phi^{\prime\prime}(x_1)
= 
\frac{\mathrm{e}}{\varepsilon_0 \, \varepsilon}
\!
\left[
2 \, c_0 \, \sinh \! \left(\!
\frac{\mathrm{e}\, \phi(x_1)}{k_b T}
\!\right)
+
\frac{N_p^2}{L_2 \, L_3}
\!
\sum_{j=-\infty}^\infty 
\!\!
\delta\big(x_1 - (j+1/2) L_1\big)
-
\delta\big(x_1 - j L_1\big)
\right]\!.
\label{Poissonx1}
\end{equation}
Since $\phi(x_1)$ is $L_1$-periodic function and symmetric around $x_1=L_1/2$, we can find it by solving equation~(\ref{Poissonx1}) in interval $[0,L_1/2]$. This is equivalent to solving
\begin{equation}
\phi^{\prime\prime}(x_1)
= 
\frac{2 \, \mathrm{e} \, c_0}{\varepsilon_0 \, \varepsilon} \, \sinh \! \left(
\frac{\mathrm{e}\, \phi(x_1)}{k_b T}
\right) 
\qquad
\mbox{in}
\quad
x \in [0,L_1/2],
\label{Poissonx1b}
\end{equation}
with boundary conditions
\begin{equation}
\phi^\prime(0)
=
- \frac{N_p^2 \, \mathrm{e}}{2 \, \varepsilon_0 \, \varepsilon \, L_2 \, L_3},
\qquad
\phi^\prime(L_1/2)
=
-
\frac{N_p^2 \, \mathrm{e}}{2 \, \varepsilon_0 \, \varepsilon \, L_2 \, L_3},
\label{boundconditionPB}
\end{equation}
and the normalization condition~(\ref{normcondition}). The solutions are presented in Figure~\ref{fig1}(b), where we plot $c^-$ and $c^+$ given by~$(\ref{Boltzmann})$ as a function of $x_1$ for three different choices of $N$ and $N_{\mathrm{p}}$, namely: (i) $N=100$ and $N_{\mathrm{p}}=6$; (ii) $N=100$ and $N_{\mathrm{p}}=4$; and (iii) $N=50$ and $N_{\mathrm{p}}=4$. In our calculations based on solving equation~(\ref{Poissonx1b}), we assume that $N/2$ of mobile ions are in interval $[0,L_1/2]$. To visualize the results in Figure~\ref{fig1}(b), we extend the calculated solution of~(\ref{Poissonx1b}) from $[0,L_1/2]$ to the whole interval $[0,L_1]$ by symmetry.

\subsection{BD implementation} 
\label{sec42}

The equilibrium solution of macroscopic equations equations given by the Poisson-Boltzmann system~(\ref{Boltzmann})--(\ref{Poissonx1}) in Figure~\ref{fig1}(b) provides the coarsest description of our model system with Na$^+$ and Cl$^-$ ions. In this section, we will present an implementation of the most detailed (microscopic, individual-based) model which is given as a BD simulation of the system of N sodium and N chloride ions in the cuboid domain~(\ref{cuboidomega}) with permanent charges at locations~(\ref{permNachargeslocations})--(\ref{permClchargeslocations}) and with periodic boundary conditions. In particular, the permanent charges are periodically repeated in all planes $x_1 = j L_1/2$, where $j$ is an integer. To simulate the evolution of the BD model, we will choose a small time step $\Delta t$ and we discretize equation~(\ref{BDSDEform3D}) by using the Euler-Maruyama method. We get
\begin{equation}
\mathbf{X}_j^\pm(t+\Delta t)
= 
\mathbf{X}_j^\pm(t)
- \alpha^\pm
\,
\nabla
\!\left(
\Phi^\pm_j \big({\mathbf{X}}_j^\pm\big)
\right)
\,
\Delta t
+ 
\sqrt{2D^{\pm} \Delta t} \,
\boldsymbol{\xi}_j\,,
\label{BDSDEform3DEM}
\end{equation}
for $j=1,2,\dots,N,$ where the coordinates of vector $\boldsymbol{\xi}_j$ are sampled from a normal distribution with zero mean and unit variance and $\Phi^\pm_j\big({\mathbf x}\big)
\equiv \Phi^\pm_j\big({\mathbf x};  \widetilde{\mathbf{X}}_j^\pm \big)$ is given by (\ref{nonsingpotentialsplus})--(\ref{nonsingpotentialsminus}). In our BD simulations based on equation~(\ref{BDSDEform3DEM}), we need to calculate the potential gradient at every time step. This includes sums~(\ref{notation2}) over all integer valued vectors~(\ref{notation1}). To evaluate these sums efficiently, we first
decompose the potential into two terms~(\ref{errordefpmj}) and rewrite the BD equation~(\ref{BDSDEform3D}) as (\ref{BDSDEform3Delpot}). The first term
$E^\pm_j\big({\mathbf X}_j^\pm;  \widetilde{\mathbf{X}}_j^\pm \big)$
only contains short-range interactions and we can calculate it by replacing the sum of infinitely many terms~(\ref{notation2}) with the most important term in this sum, which corresponds
to the copy of the point closest to ${\mathbf X}_j^\pm$, i.e. we use the minimum--image convention to approximate all short-range interactions in $E^\pm_j\big({\mathbf X}_j^\pm;  \widetilde{\mathbf{X}}_j^\pm \big)$. Moreover, we denote the Euclidean distance $\parallel \!\cdot \!\parallel$ modified by the minimum-image convention by
\begin{equation}
|{\mathbf{z}}_1-{\mathbf{z}}_2|_{{\mathrm{min}}}
=
\min_{\boldsymbol{\ell}}
\parallel\! 
{\mathbf{z}}_1 - {\mathbf{z}}_j - {\mathbf{L}}_{\boldsymbol{\ell}} \!\parallel
\qquad
\mbox{for any}
\quad
{\mathbf{z}}_1, \, {\mathbf{z}}_2 \in {\mathbb R}^3.
\label{minimimage}
\end{equation} 
The second term in~(\ref{errordefpmj}), 
$q^\pm \, \widehat{\phi}^\pm_j\big(\textbf{X}_j^\pm;  \widetilde{\mathbf{X}}_j^\pm\big)$, can be expressed using $\widehat{\phi}\big(\mathbf{x}; \overline{\mathbf{y}}\big)$ that solves the Poisson equation~(\ref{potentialxovery}). We divide it into two terms corresponding to the mobile and fixed, ions, respectively. We obtain
$$
\widehat{\phi}\big(\mathbf{x}; \overline{\mathbf{y}}\big)
=
\widehat{\phi}_{\mathrm{m}} \big(\mathbf{x}; \overline{\mathbf{y}}\big)
+
\widehat{\phi}_\mathrm{p} 
\big(\mathbf{x}\big), 
$$
where 
$\widehat{\phi}_{\mathrm{m}} \big(\mathbf{x}; \overline{\mathbf{y}}\big)$
and $\widehat{\phi}_\mathrm{p} 
\big(\mathbf{x}\big)$
solve the Poisson equations in the following forms
\begin{equation}
\; \nabla^2 \widehat{\phi}_{\mathrm{m}}\big(\mathbf{x}; \overline{\mathbf{y}}\big) 
= -\frac{1}{\varepsilon_0 \, \varepsilon} \sum_{\boldsymbol{\ell}}
\left[ 
\sum_{i=1}^{N}
\Big(
q^+ \,
\delta^3
\big(\mathbf{x}\!-\!\mathbf{y}_i^+
\!\!-\! {\mathbf{L}}_{\boldsymbol{\ell}}\big) 
+ 
q^- \,
\delta^3
\big(\mathbf{x}\!-\!\mathbf{y}_i^-
\!\!-\! {\mathbf{L}}_{\boldsymbol{\ell}}\big)\!
\Big)\!
\right]
\label{potentialmobile}
\end{equation}
and
\begin{equation}
\; \nabla^2 \widehat{\phi}_{\mathrm{p}}\big(\mathbf{x}\big) 
= -\frac{1}{\varepsilon_0 \, \varepsilon} \sum_{\boldsymbol{\ell}}
\varrho_{\mathrm{p}}(\mathbf{x}\!- \!{\mathbf{L}}_{\boldsymbol{\ell}}).
\label{potentialpermanent}
\end{equation}
To solve~(\ref{potentialmobile}), we will use the Ewald summation~\cite{Ewald:1921:BOE,Frenkel:2002:UMS}. We get
\begin{eqnarray}
\widehat{\phi}_{\mathrm{m}}\big(\mathbf{x}; \overline{\mathbf{y}}\big)
&=&
\frac{1}{4 \pi \varepsilon_0 \, \varepsilon}\sum_{i=1}^{N}
\left(
q^+ 
\frac{\text{erfc}(\beta |\mathbf{x}\!-\!\mathbf{y}_i^+|_{\mathrm{min}})}{|\mathbf{x}\!-\!\mathbf{y}_i^+|_{\mathrm{min}}}
+ 
q^- 
\frac{\text{erfc}(\beta |\mathbf{x}\!-\!\mathbf{y}_i^-|_{\mathrm{min}})}{|\mathbf{x}\!-\!\mathbf{y}_i^-|_{\mathrm{min}}}
\right) \nonumber
\\
&+&
\frac{1}{\varepsilon_0 \, \varepsilon \, L_1  L_2 L_3} \sum_{\boldsymbol{\ell} \ne [0,0,0]}
\frac{1}{|{\mathbf{k}}_{\boldsymbol{\ell}}|^2}
\, \exp 
\!\left(
- \frac{|{\mathbf{k}}_{\boldsymbol{\ell}}|^2}{4 \beta^2}
\right)
\label{ewaldsum}
\\
&& \qquad \qquad \qquad
\times
\sum_{i=1}^{N}
\left(
q^+ 
\exp\big(\mathrm{i} \, {\mathbf{k}}_{\boldsymbol{\ell}} \cdot
(\mathbf{x}\!-\!\mathbf{y}_i^+)\big)
+ 
q^- 
\exp\big(\mathrm{i} \, {\mathbf{k}}_{\boldsymbol{\ell}}\cdot
(\mathbf{x}\!-\!\mathbf{y}_i^-)\big)
\right)
\nonumber
\end{eqnarray}
where $\beta>0$ is a parameter, $|\cdot|_{\mathrm{min}}$ denotes the distance modified by the minimum image convention~(\ref{minimimage}) and
\begin{equation}
{\mathbf{k}}_{\boldsymbol{\ell}}
=
\left[
\frac{2 \pi \ell_1}{L_1},
\, \frac{2 \pi \ell_2}{L_2},
\, \frac{2 \pi \ell_3}{L_3}
\right]
\qquad
\mbox{for any integer valued vector}
\quad
\boldsymbol{\ell}
=
[\ell_1,\ell_2,\ell_3].
\label{notation3}
\end{equation}
 Since the permanent charge density is given by~(\ref{varrhopexample}), the equation~(\ref{potentialpermanent}) has the same functional form as the equation~(\ref{potentialmobile}) with permanent charges replacing the mobile charges. In particular, the potential $\widehat{\phi}_\mathrm{p} 
\big(\mathbf{x}\big)$ can be calculated by the Ewald summation formula~(\ref{ewaldsum}) with permanent charges replacing the mobile charges, with some computations performed only once at the beginning of the simulation. We define
\begin{equation}
\quad
\theta_{\mathrm{p}}(\boldsymbol{\ell})
=
\frac{\exp 
\!\left(\!
- |{\mathbf{k}}_{\boldsymbol{\ell}}|^2/(4 \beta^2)
\right)}{|{\mathbf{k}}_{\boldsymbol{\ell}}|^2 \, \varepsilon_0 \, \varepsilon \, L_1  L_2 L_3}
\sum_{m=1}^{N_{\mathrm{p}}}
\!
\sum_{n=1}^{N_{\mathrm{p}}}
q^+
\exp \!
\big(\mathrm{i} \, {\mathbf{k}}_{\boldsymbol{\ell}} {\hskip 0.3mm} \cdot {\hskip 0.3mm} \mathbf{z}_{m,n}^+
\big)
\,+\, 
q^- 
\exp \! \big(\mathrm{i} \, {\mathbf{k}}_{\boldsymbol{\ell}} {\hskip 0.3mm} \cdot {\hskip 0.3mm} \mathbf{z}_{m,n}^-
\big),
\label{thetalsumsperm}
\end{equation}
where $\mathbf{z}_{m,n}^+$ and $\mathbf{z}_{m,n}^-$ are positions of permanent positive and negative charges given by~(\ref{permNachargeslocations}) and~(\ref{permClchargeslocations}), respectively.
Our BD algorithm uses equation~(\ref{BDSDEform3DEM}) at every time step, which requires calculating the gradient of~(\ref{ewaldsum}). Let $r$ be the distance between two ions with separation vector ${\mathbf r}$, i.e. $r=|{\mathbf r}|$. Differentiating~(\ref{eqpotential}) and the first term (real part) of the potential~(\ref{ewaldsum}), we define (the real part of) the force term between the ions as
\begin{equation}
{\mathbf F}_k(\mathbf r)
=
\left(
\frac{12 \, A_k}{r^{14}}
-
\frac{6 \, B_k}{r^8}
+
\frac{2 \, \beta}{\sqrt{\pi}}
\frac{C_k \, \exp(-\beta^2 r^2)}{r^2}
+
\frac{C_k \, \text{erfc}(\beta r)}{r^3}
\right) {\mathbf r}
\,,
\label{fkrforce}
\end{equation}
where $k=1$, $k=2$, and $k=3$, correspond to interactions between two Na$^+$ ions, one Na$^+$ and one Cl$^-$ ion, and two Cl$^-$ ions, respectively. 

One iteration of the BD algorithm ({\it i.e.} the update of the system from time $t$ to time $t+\Delta t$) is described in Table~\ref{table2} as Algorithm~[A1]--[A6]. We denote the drift term $- \alpha^\pm
\,
\nabla
\!\left(
\Phi^\pm_j \big({\mathbf{X}}_j^\pm\big)
\right)$ in equation~(\ref{BDSDEform3DEM}) as ${\mathbf{a}}_j^\pm(t)$, i.e. we rewrite the discretized equation~(\ref{BDSDEform3DEM}) as
\begin{equation}
\mathbf{X}_j^\pm(t+\Delta t)
= 
\mathbf{X}_j^\pm(t)
+
{\mathbf{a}}_j^\pm(t)
\,
\Delta t
+ 
\sqrt{2D^{\pm} \Delta t} \,
\xi_j\,,
\label{BDSDEform3DEMwithalpha}
\end{equation}
for $j=1,2,\dots,N,$ where drift term ${\mathbf{a}}_j^\pm(t)$ is calculated in steps [A1]--[A5]. In step~[A1], we initialize the drift term using the components of forces corresponding to the permanent charges, putting
\begin{eqnarray}
{\mathbf{a}}_j^+(t) & := & \alpha^+ 
\sum_{m=1}^{N_{\mathrm{p}}}
\sum_{n=1}^{N_{\mathrm{p}}}
{\mathbf F}_1
\big(
({\textbf X}_j^+ - \mathbf{z}_{m,n}^+)_{\mathrm{min}} 
\big) 
+
{\mathbf F}_2
\big(
({\textbf X}_j^+ - \mathbf{z}_{m,n}^-)_{\mathrm{min}} 
\big) 
\label{permforcep}
\\
{\mathbf{a}}_j^-(t) & := & \alpha^-
\sum_{m=1}^{N_{\mathrm{p}}}
\sum_{n=1}^{N_{\mathrm{p}}}
{\mathbf F}_2
\big(
({\textbf X}_j^- - \mathbf{z}_{m,n}^+)_{\mathrm{min}} 
\big) 
+
{\mathbf F}_3
\big(
({\textbf X}_j^- - \mathbf{z}_{m,n}^-)_{\mathrm{min}} 
\big) 
\label{permforcen}
\end{eqnarray}
where we substitute separation vectors modified by the minimum image convention~(\ref{minimimage}) in the formula~(\ref{fkrforce}), using notation 
$(\cdot)_{\mathrm{min}}$ and $|\cdot|_{\mathrm{min}}$ to denote the corresponding modified vectors and norms, respectively. In step [A2], we add force terms corresponding to interactions between mobile Na$^+$ ions. This is followed in steps [A3] and [A4] by adding force terms corresponding to Na$^+$--Cl$^-$ and Cl$^-$--Cl$^-$ interactions, respectively, between mobile ions. In step [A5], we calculate sum
\begin{equation}
\quad
\theta(\boldsymbol{\ell})
=
\theta_{\mathrm{p}}(\boldsymbol{\ell})
+
\frac{\exp 
\!\left(\!
- |{\mathbf{k}}_{\boldsymbol{\ell}}|^2/(4 \beta^2)
\right)}{|{\mathbf{k}}_{\boldsymbol{\ell}}|^2 \, \varepsilon_0 \, \varepsilon \, L_1  L_2 L_3}
\sum_{i=1}^{N}
q^+ 
\,
\exp
\big(\mathrm{i} \, {\mathbf{k}}_{\boldsymbol{\ell}} \cdot {\textbf X}_i^+
\big)
+ 
q^- 
\exp\big(\mathrm{i} \, {\mathbf{k}}_{\boldsymbol{\ell}}\cdot
{\textbf X}_i^-
\big),
\label{thetalsums}
\end{equation}
which is needed to evaluate the Fourier part of the potential~(\ref{ewaldsum}). It includes the precomputed term $\theta_{\mathrm{p}}(\boldsymbol{\ell})$, given by~(\ref{thetalsumsperm}), corresponding to the permanent charges. The summation in step [A5] is over all $\boldsymbol{\ell} \in {\mathcal L}(\kappa)$, where $\kappa \in {\mathbb N}$ is a parameter to be specified, and set ${\mathcal L}(\kappa)$ is defined by
\begin{equation}
{\mathcal L}(\kappa)
=
\left\{
\boldsymbol{\ell}
= [\ell_1,\ell_2,\ell_3]
\; \Big| \;
|\ell_1| \le \kappa,
\,
|\ell_2| \le \kappa,
\,
|\ell_3| \le \kappa
\;
\mbox{and}
\;
\boldsymbol{\ell} \ne [0,0,0]
\right\}.
\label{truncation}
\end{equation}
Finally, after we evaluate the drift term in steps [A1]--[A5], we use equation~(\ref{BDSDEform3DEMwithalpha}) to calculate the positions of mobile ions at time $t+\Delta t$ in step [A6].

\begin{table}
\boxed{\hbox{\hskip 1mm\hsize=0.965\hsize\vbox{\vskip 0.5mm
\parindent -8.5mm \leftskip 8.5mm
[A1] Initialize ${\mathbf{a}}_j^\pm(t)$, for $j=1,2,\dots,N,$ using formulas~(\ref{permforcep})--(\ref{permforcen}).
\par \vskip 1mm
[A2] For each pair of Na$^+$ ions, labelled $i$ and $j$ (for $i \ne j$), calculate force ${\mathbf F}_1(\mathbf r)$ between them by~(\ref{fkrforce}), where ${\mathbf r} = ({\textbf X}_i^+ - {\textbf X}_j^+)_{\mathrm{min}}.$ Put \hfill\break
\rule{0pt}{1pt} \hskip 1cm ${\mathbf{a}}_i^+(t) := {\mathbf{a}}_i^+(t)
+ \alpha^+ {\mathbf F}_1\big(({\textbf X}_i^+ - {\textbf X}_j^+)_{\mathrm{min}}\big)$ \hfill\break
\rule{0pt}{1pt} \hskip 1cm ${\mathbf{a}}_j^+(t) := {\mathbf{a}}_j^+(t)
- \alpha^+ {\mathbf F}_1\big(({\textbf X}_i^+ - {\textbf X}_j^+)_{\mathrm{min}}\big)$ \par \vskip 1mm
[A3] For each Na$^+$ ion, labelled $i$, and each Cl$^-$ ion, labelled $j$, calculate force ${\mathbf F}_2(\mathbf r)$ between them by~(\ref{fkrforce}), where ${\mathbf r} = ({\textbf X}_i^+ - {\textbf X}_j^-)_{\mathrm{min}}.$ Put \hfill\break
\rule{0pt}{1pt} \hskip 1cm ${\mathbf{a}}_i^+(t) := {\mathbf{a}}_i^+(t)
+ \alpha^+ {\mathbf F}_2\big(({\textbf X}_i^+ - {\textbf X}_j^-)_{\mathrm{min}}\big)$ \hfill\break
\rule{0pt}{1pt} \hskip 1cm ${\mathbf{a}}_j^-(t) := {\mathbf{a}}_j^-(t)
- \alpha^- {\mathbf F}_2\big(({\textbf X}_i^+ - {\textbf X}_j^-)_{\mathrm{min}}\big)$
\par \vskip 1mm
[A4] For each pair of Cl$^-$ ions, labelled $i$ and $j$ (for $i \ne j$), calculate force ${\mathbf F}_3(\mathbf r)$ between them by~(\ref{fkrforce}), where ${\mathbf r} = ({\textbf X}_i^- - {\textbf X}_j^-)_{\mathrm{min}}.$ Put \hfill\break
\rule{0pt}{1pt} \hskip 1cm ${\mathbf{a}}_i^-(t) := {\mathbf{a}}_i^-(t)
+ \alpha^- {\mathbf F}_3\big(({\textbf X}_i^- - {\textbf X}_j^-)_{\mathrm{min}}\big)$ \hfill\break
\rule{0pt}{1pt} \hskip 1cm ${\mathbf{a}}_j^-(t) := {\mathbf{a}}_j^-(t)
- \alpha^- {\mathbf F}_3\big(({\textbf X}_i^- - {\textbf X}_j^-)_{\mathrm{min}}\big)$ 
\par \vskip 1mm
[A5] Calculate sum~(\ref{thetalsums}) for all $\boldsymbol{\ell} \in {\mathcal L}(\kappa)$, given by~(\ref{truncation}), and put
\hfill\break
\rule{0pt}{1pt} \hskip 1cm ${\mathbf{a}}_j^+(t) := {\mathbf{a}}_j^+(t)
+ \displaystyle  
\alpha^+ q^+
\sum_{\boldsymbol{\ell} \in {\mathcal L}(\kappa)}
\mbox{Im} \Big(
\theta(\boldsymbol{\ell}) 
\exp\big(\mathrm{i} {\mathbf{k}}_{\boldsymbol{\ell}} \cdot {\textbf X}_j^+
\big) \!\Big) \, {\mathbf{k}}_{\boldsymbol{\ell}}
$ \hfill\break
\rule{0pt}{1pt} \hskip 1cm ${\mathbf{a}}_j^-(t) := {\mathbf{a}}_j^-(t) \displaystyle
+ \alpha^- q^- \sum_{\boldsymbol{\ell} \in {\mathcal L}(\kappa)} \mbox{Im} \Big(
\theta(\boldsymbol{\ell}) 
\exp\big(\mathrm{i} \, {\mathbf{k}}_{\boldsymbol{\ell}} \cdot {\textbf X}_j^-
\big)\!\Big) \, {\mathbf{k}}_{\boldsymbol{\ell}}
$ \hfill\break
where $\mbox{Im}(\cdot)$ denotes the imaginary part of a complex number.
\par 
\vskip 1mm
[A6] Generate $6N$ coordinates of vectors $\boldsymbol{\xi}_j$, for $j=1,2,\dots,N$, as normally distributed numbers with zero mean and unit variance. Calculate the positions of mobile ions at time $t+\Delta t$ by using equation~(\ref{BDSDEform3DEMwithalpha}).
\par \vskip 0.5mm}\hskip 1mm}}
\vskip 1mm
\caption{\label{table2}
One iteration of the BD algorithm to calculate the positions of ions at time $t+\Delta t$ given positions of ions at time $t$.}
\end{table}

\begin{figure}[t]
\epsfig{file=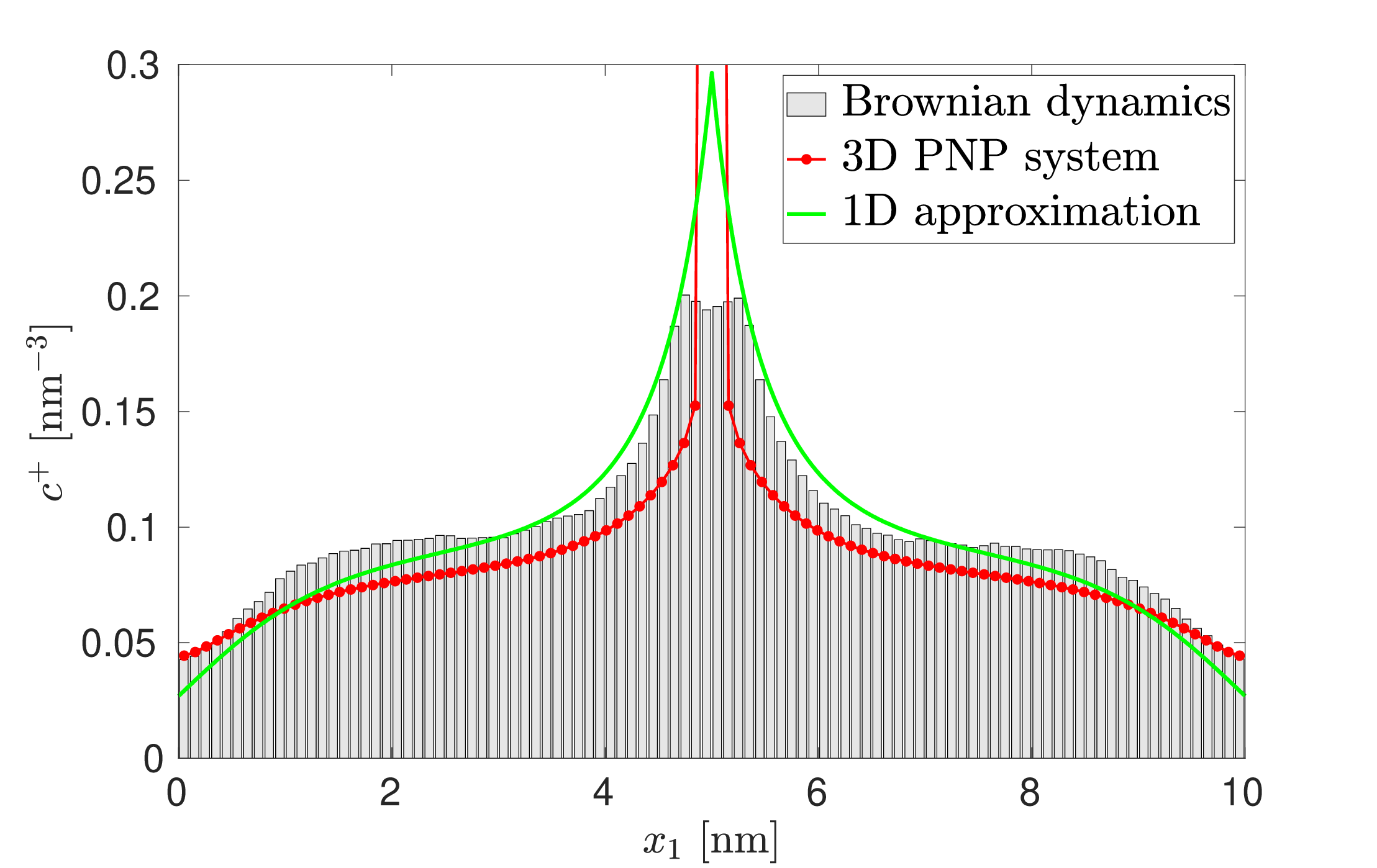,height=3.94cm}
\hskip 0.1mm 
\epsfig{file=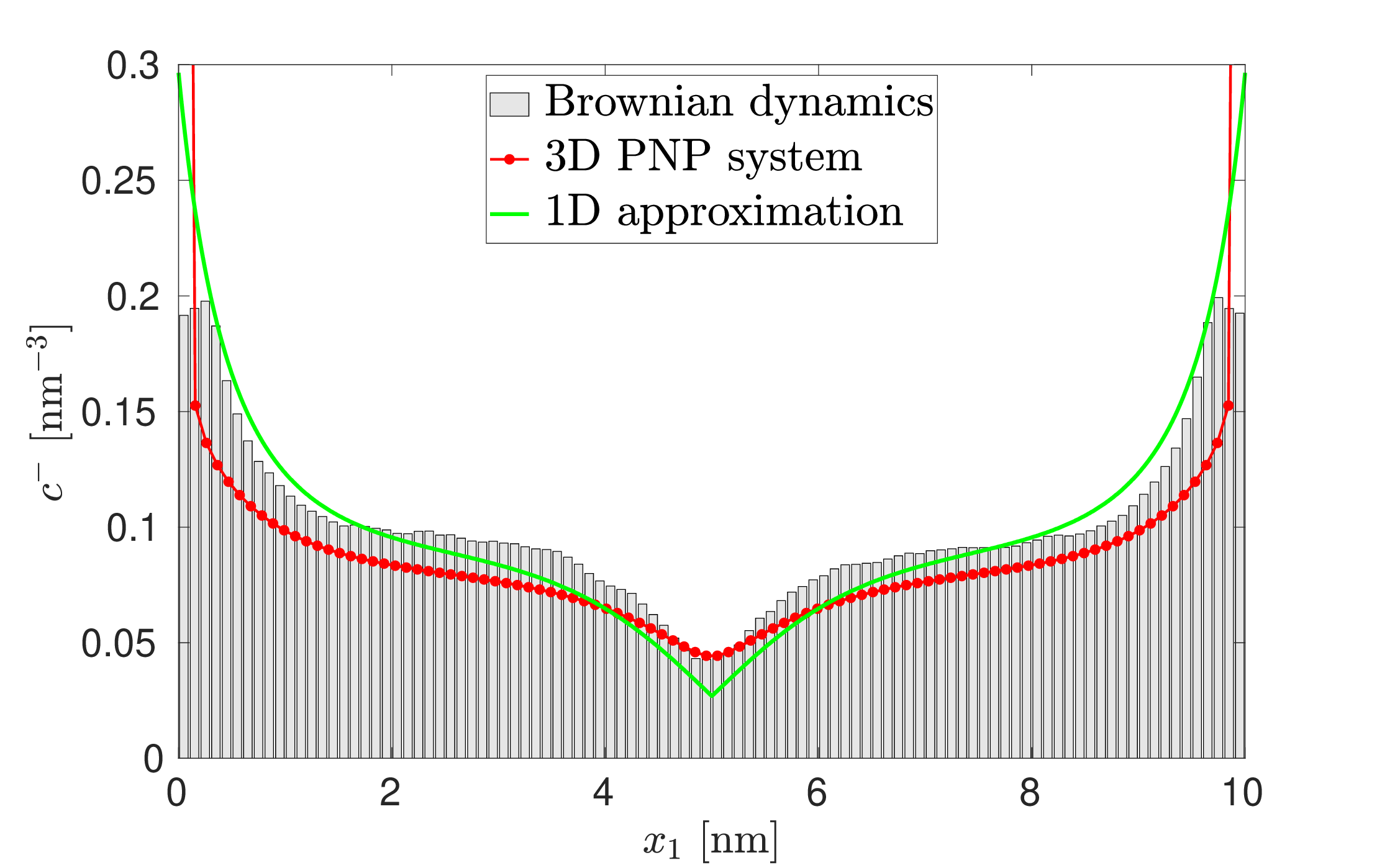,height=3.94cm}
\vskip -4cm
(a) \hskip 5.8cm (b)
\vskip 3.45cm
\caption{The equilibrium distributions calculated using Algorithm~{\rm [A1]--[A6]} for $N=100$, $N_p=6$, $\kappa=3$ and $\beta=0.05$ are visualized as gray histograms for {\rm (a)} mobile {\rm Na}$^+$ ions;  and {\rm (b)} mobile Cl$^-$ ions. The results are compared with density $c^\pm$ calculated for the same parameter values in Figure~$\ref{fig1}(b)$ by using a one-dimensional approximation of the PNP system, given by $(\ref{Poissonx1})$ with boundary conditions~$(\ref{boundconditionPB})$ $($green lines$)$. The red lines are calculated using Algorithm~{\rm [B1]--[B4]} for numerically solving the three-dimensional {\rm PNP} system~$(\ref{NernstPlanck})$--$(\ref{Poisson})$.}
\label{fig2}
\end{figure}%

In Figure~\ref{fig2}, we present illustrative results calculated by Algorithm~[A1]--[A6]. We use $N=100$ mobile Na$^+$ ions and $N=100$ mobile Cl$^+$ ions in domain $\Omega$ given by~(\ref{cuboidomega}) with $L_1 = L_2 = L_3 = 10 \, \mbox{nm}$ and the permanent charges at locations~(\ref{permNachargeslocations})--(\ref{permClchargeslocations}) with $N_{\mathrm{p}}=6$. To apply our BD Algorithm~[A1]--[A6], we need to specify parameters $\beta>0$ and $\kappa \in {\mathbb N}$, which are used to implement the periodic boundary conditions using the Ewald summation. In our illustrative simulations, we choose $\kappa = 3$ in (\ref{truncation}), which means that we evaluate $|{\mathcal L}(\kappa)| = (2\kappa+1)^3 - 1 = 342$ terms in the sum in step [A5]. This truncation of the infinite sum introduces an error in the evaluation of the Fourier part of the Ewald sum. The choice of the parameter $\beta>0$ can then be optimized in the way that the error of the real part of the Ewald sum (which dominates for small values of $\beta$) is comparable to the error of the Fourier part (which dominates for large values of $\beta$). Using $\kappa=3,$ we find the optimal value of $\beta$ to be $\beta=0.05,$ which is used to obtain the results presented in Figure~\ref{fig2}, where we visualize the spatial histogram (density of ions) obtained by running the long-time simulation of Algorithm~[A1]--[A6] over $10^7$ time steps of length $\Delta t = 10^{-2} \, \mbox{ps}.$ To enable direct comparison, with our one-dimensional results in Figure~\ref{fig1}(b), we plot the calculated equilibrium density over the first coordinate, $x_1$, integrating over the $x_2$ and $x_3$ coordinates. We present this one-dimensional (marginal) distribution of mobile Na$^+$ ions in Figure~\ref{fig2}(a) as the green histogram. We compare it with the results (blue line) calculated by the approximate one-dimensional theory developed in Section~\ref{sec41}. We observe similar qualitative behaviour, but the results quantitatively differ. We confirm this observation in Figure~\ref{fig2}(b) where we present the BD~results for mobile Cl$-$ ions as the green histogram. There are a number of reasons behind the observed difference. One of them is that our solution to the PNP system in Section~\ref{sec41} is itself a one-dimensional approximation. We will next compare it with numerically solving the PNP system~(\ref{NernstPlanck})--(\ref{Poisson}) in the three-dimensional domain. This numerical approach will also be useful for developing the multi-resolution method in Section~\ref{sec5}.

\subsection{Numerical method for solving macroscopic PNP equations}

\label{sec43}

To solve the macroscopic PNP equations~(\ref{NernstPlanck})--(\ref{Poisson}) in the three-dimensional cuboid domain~(\ref{cuboidomega}), we use a cuboid mesh with $n_1 \times n_2 \times n_3$ meshpoints, where $n_i \in {\mathbb N}$ for $i=1,2,3$ and define
\begin{equation}
\Delta x_1 =  \frac{L_1}{n_1}\,,
\qquad
\Delta x_2 =  \frac{L_2}{n_2}
\qquad
\mbox{and}
\qquad
\Delta x_3 = \frac{L_3}{n_3}\,.
\label{gridsizes}
\end{equation}
The meshpoints are at locations
\begin{equation}
\mathbf{x}_{i,j,k}
=
\big[
(i-1/2) \, \Delta x_1,
\,
(j-1/2) \, \Delta x_2,
\,
(k-1/2) \, \Delta x_3
\big]
\label{meshpoints}
\end{equation}
for $i=1, 2, \dots, n_1,$ $j=1, 2, \dots, n_2,$ $k=1, 2, \dots, n_3,$ and we denote
\begin{equation}
c^\pm_{i,j,k}(t)
=
c^\pm(\mathbf{x}_{i,j,k},t),
\qquad
\phi_{i,j,k}(t)
=
\phi(\mathbf{x}_{i,j,k},t).
\label{cijkphiijkdef}
\end{equation}
Using finite differences to discretize the right-hand side of equation~(\ref{NernstPlanck}), we obtain
\begin{equation}
\frac{\mbox{d} c^\pm_{i,j,k}}{\mbox{d} t}
=
{\mathcal A}_{i,j,k} (c^\pm,\phi)
\label{NernstPlanckdiscr}
\end{equation}
where ${\mathcal A}_{i,j,k} (c^\pm,\phi)$ is defined by
\begin{eqnarray*}
&& {\mathcal A}_{i,j,k} (c^\pm,\phi)
=
D^\pm
\left( \frac{c^\pm_{i+1,j,k} - 2 c^\pm_{i,j,k} + c^\pm_{i-1,j,k}}{\Delta x_1^2}
+
\frac{c^\pm_{i,j+1,k} - 2 c^\pm_{i,j,k} + c^\pm_{i,j-1,k}}{\Delta x_2^2}
\right.
\\
&& \left. 
\hskip 4cm +
\frac{c^\pm_{i,j,k+1} - 2 c^\pm_{i,j,k} + c^\pm_{i,j,k-1}}{\Delta x_3^2}
\right)
\\
&& + \,
\frac{D^\pm \, q^\pm}{2 k_b T}
\left(
\frac{
\big(c^\pm_{i+1,j,k} + c^\pm_{i,j,k}\big) (\phi_{i+1,j,k} - \phi_{i,j,k})
+
\big(c^\pm_{i-1,j,k} + c^\pm_{i,j,k}\big) (\phi_{i-1,j,k} - \phi_{i,j,k})
}{\Delta x_1^2}
\right.
\\
&&\hskip 1.5cm
+ \, 
\frac{
\big(c^\pm_{i,j+1,k} + c^\pm_{i,j,k}\big) (\phi_{1,j+1,k} - \phi_{i,j,k})
+
\big(c^\pm_{i,j-1,k} + c^\pm_{i,j,k}\big) (\phi_{i,j-1,k} - \phi_{i,j,k})
}{\Delta x_2^2}
\\
&&\hskip 1.2cm
+ \, \left. 
\frac{
\big(c^\pm_{i,j,k+1} + c^\pm_{i,j,k}\big) (\phi_{i,j,k+1} - \phi_{i,j,k})
+
\big(c^\pm_{i,j,k-1} + c^\pm_{i,j,k}\big) (\phi_{i,j,k-1} - \phi_{i,j,k})
}{\Delta x_3^2}
\right),
\end{eqnarray*}
with the convention that indices $i,$ $j,$ $k$ are periodic with periods $n_1$, $n_2$ and $n_3$, respectively. For example, $c^+_{n_1+1,j,k}$ is identified with $c^+_{1,j,k}$ to implement the periodic boundary conditions. One iteration of our algorithm for solving the PNP equations~(\ref{NernstPlanck})--(\ref{Poisson}) is given in Table~\ref{table3} as Algorithm~[B1]--[B4]. It calculates concentrations $c^\pm_{i,j,k}(t+\delta t)$ and potential $\phi_{i,j,k}(t + \delta t)$ at time $t + \delta t$ given the concentrations $c^\pm_{i,j,k}(t)$ and potential $\phi_{i,j,k}(t)$ at time $t$, where $\delta t$ is the time step used for solving the PNP equations~(\ref{NernstPlanck})--(\ref{Poisson}). We note that we use different notation to denote the time step, $\Delta t$, in our BD~model in Table~\ref{table2}, and the time step $\delta t$ to solve the PNP equations~(\ref{NernstPlanck})--(\ref{Poisson}) in Table~\ref{table3}, because time steps $\delta t$ and $\Delta t$ can be, in general, different. 

In step [B1], we solve the system of ODEs~(\ref{NernstPlanckdiscr}) over the time interval $[t,t+\delta t).$ Assuming that $\delta t$ is chosen sufficiently small, we can update the system using the forward Euler scheme
\begin{eqnarray}
c^+_{i,j,k}(t+\delta t)
&:=&
c^+_{i,j,k}(t)
\,
+
\, \delta t \,
{\mathcal A}_{i,j,k} \big(c^+(t),\phi(t)\big) \, .
\label{eulera}
\\
c^-_{i,j,k}(t+\delta t)
&:=&
c^-_{i,j,k}(t)
\,
+
\, \delta t \,
{\mathcal A}_{i,j,k} \big(c^-(t),\phi(t)\big) \, .
\label{eulerb}
\end{eqnarray}
In steps [B2]--[B4], we calculate the potential $\phi_{i,j,k}(t+\delta t)$ by solving the Poisson equation~(\ref{Poisson}). We write it as
\begin{equation}
\phi_{i,j,k}(t+\delta t)
:=
\phi^m_{i,j,k}
+
\widehat{\phi}_{\mathrm{p}}\big(\mathbf{x}_{i,j,k}\big) \,,
\label{phiijkdec}
\end{equation}
where $\widehat{\phi}_{\mathrm{p}}\big(\mathbf{x}\big)$  is the solution of (\ref{potentialpermanent}) giving the potential corresponding to the permanent charges. This potential is the same as in our BD~simulations in Section~\ref{sec42} and is precomputed at the beginning of the simulation. Subtracting equations~(\ref{Poisson}) and (\ref{potentialpermanent}), we obtain $\phi^m_{i,j,k}$ in equation~(\ref{phiijkdec}) as $\phi^m_{i,j,k}
= \phi^m ({\mathbf x}_{i,j,k})$, where $\phi^m$ solves the Poisson equation
\begin{equation}
\nabla^2 \phi^m
= -\frac{1}{\varepsilon_0 \, \varepsilon} \left[
q^+ \, c^+ 
\,+\, 
q^- c^-\right].
\label{potc}    
\end{equation}
To solve equation~(\ref{potc}), we use the central difference method to discretize the Laplacian to obtain
\begin{eqnarray}
&&
\frac{\phi^m_{i+1,j,k} - 2 \phi^m_{i,j,k} + \phi^m_{i-1,j,k}}{\Delta x_1^2}
\;+\;
\frac{\phi^m_{i,j+1,k} - 2 \phi^m_{i,j,k} + \phi^m_{i,j-1,k}}{\Delta x_2^2}
\nonumber
\\
&& 
\hskip 1cm + \;
\frac{\phi^m_{i,j,k+1} - 2 \phi^m_{i,j,k} + \phi^m_{i,j,k-1}}{\Delta x_3^2}
\; = \;
-\frac{1}{\varepsilon_0 \, \varepsilon} \left[
q^+ \, c^+_{i,j,k}
\,+\, 
q^- c^-_{i,j,k}
\right]
\label{potcdisc}
\end{eqnarray}
and apply the discrete Fourier transform. In step~[B2], we use the fast Fourier transform algorithm to calculate the discrete Fourier transform of the right-hand side of equation~(\ref{potcdisc}), which is used in step~[B3] to calculate $\phi^m_{i,j,k}$ by using the inverse fast Fourier transform algorithm. The corresponding factor $\lambda(i^\prime,j^\prime,k^\prime)$ is obtained by calculating the discrete Fourier transform of the discretized Laplacian on the left hand side of equation~(\ref{potcdisc})
as
\begin{eqnarray}
\lambda(i^\prime,j^\prime,k^\prime)
&=&
\frac{1}{2}
\left[
\left(
1-\cos \frac{2\pi(i^\prime\!-\!1)}{n_1} 
\right) 
\frac{n_1^2}{L_1^2}
+
\left(
1-\cos \frac{2\pi(j^\prime\!-\!1)}{n_2} 
\right) 
\frac{n_2^2}{L_2^2}
\nonumber
\right.
\\
&&
\left.
\quad +
\left(
1-\cos \frac{2\pi(k^\prime\!-\!1)}{n_3} 
\right) 
\frac{n_3^2}{L_3^2}
\right]^{-1}.
\label{lambdacoef}
\end{eqnarray}
The potential $\phi_{i,j,k}(t+\delta t)$ is then calculated in step~[B4] by adding the potential corresponding to the permanent charges using equation~(\ref{phiijkdec}).

\begin{table}
\boxed{\hbox{\hskip 1mm\hsize=0.955\hsize\vbox{\vskip 0.5mm
\parindent -8mm \leftskip 8mm
[B1] Calculate concentrations $c^+_{i,j,k}(t+\delta t)$ and $c^-_{i,j,k}(t+\delta t)$ by using~(\ref{eulera})--(\ref{eulerb}).
\par \vskip 1mm
[B2] Calculate the discrete Fourier transform $f_{i^\prime,j^\prime,k^\prime}$ of 
$$
-\frac{1}{\varepsilon_0 \, \varepsilon} \left[
q^+ \, c^+_{i,j,k}(t+\delta t) 
\,+\, 
q^- c^-_{i,j,k}(t+\delta t)\right]
$$
using the fast Fourier transform algorithm. \par \vskip 1mm
[B3] Calculate $\phi^m_{i,j,k}$ as the inverse discrete Fourier transform of
$
\lambda(i^\prime,j^\prime,k^\prime)   
f_{i^\prime,j^\prime,k^\prime}
$,
where $\lambda(i^\prime,j^\prime,k^\prime)$ is given by~(\ref{lambdacoef}), using the inverse fast Fourier transform algorithm.
\par \vskip 1mm
[B4] Calculate $\phi_{i,j,k}(t+\delta t)$ using equation~(\ref{phiijkdec}). 
\vskip 0.5mm}
\hskip 1mm}}
\vskip 1mm
\caption{\label{table3}
One iteration of the algorithm to solve {\rm PNP} equations~$(\ref{NernstPlanck})$--$(\ref{Poisson})$, calculating concentrations $c^\pm_{i,j,k}(t+\delta t)$ and potential $\phi_{i,j,k}(t + \delta t)$ at time $t + \delta t$ given the concentrations $c^\pm_{i,j,k}(t)$ and potential $\phi_{i,j,k}(t)$ at time $t$.}
\end{table}

In Figure~\ref{fig2}, we present illustrative results calculated using Algorithm~[B1]--[B4] as the red lines. We use $n_1 = n_2 = n_3 = 96 = 3 \times 2^5$ mesh points in each direction. The small prime factors enable a relatively simple implementation of the Cooley and Tukey algorithm~\cite{Cooley:1965:AMC} for the discrete Fourier transform. In Figure~\ref{fig2}, we present the calculated equilibrium densities as functions of the first coordinate, $x_1$, integrating over the $x_2$ and $x_3$ coordinates. To highlight the discreteness of the calculated results, we plot this one-dimensional (marginal) distribution of mobile Na$^+$ ions in Figure~\ref{fig2}(a) as the red dots (obtained at $n_1=96$ values) connected by the red line. The  one-dimensional (marginal) distribution of mobile Cl$^-$ ions is presented in Figure~\ref{fig2}(b). The calculated values at mesh points in the vicinity of the layers of permanent charges (at $x_1=0$ and $x_1=L/2$) are well above the BD results. This error cannot be improved by increasing the mesh sizes $n_1,$ $n_2$ and $n_3$, because it is caused by the error term~(\ref{errorforce}), which is missing in the PNP system. This error term includes the Lennard-Jones potential of permanent charges, which would prevent the density $c^\pm$ accumulating in the mesh points close to the permanent charges, if it was included in macroscopic description. While this could decrease the observed difference between the PNP solution and BD simulations around $x_1=0$ and $x_1=L/2$ in Figure~\ref{fig2}, it is not straightforward to add the Lennard-Jones potential to the PDE simulations, because it introduces relatively steep potential gradients in a few mesh points next to the permanent charges, leading to numerical instabilities. An alternative approach is to solve the PDEs only in sub-regions not containing permanent charges. This leads to the multi-resolution algorithm developed in Section~\ref{sec5}, where detailed BD simulations are used to capture the system dynamics in sub-regions close to the permanent charges.

\section{Multi-resolution simulations} 
\label{sec5}

Considering reaction-diffusion processes of electroneutral particles, PDE-assisted Brownian Dynamics~\cite{Franz:2013:MRA} combines BD simulations based on equation~(\ref{eq0}) with solving macroscopic reaction-diffusion PDEs in parts of the computational domain. In our model system, we would like to apply BD close to the layers of permanent charges and use PNP~equations in the rest of the computational domain. To achieve this, we need to extend the multi-resolutions approach to the models of charged particles, where the macroscopic system is given by the PNP equations~(\ref{NernstPlanck})--(\ref{Poisson}). Following~\cite{Franz:2013:MRA}, we divide our computational domain~(\ref{cuboidomega}) into two overlapping subdomains
\begin{eqnarray}
\hskip 1cm
\Omega_B
&=& 
\left( 
\Big[0,\omega_B\Big] \bigcup \left[\frac{L_1}{2} - \omega_B,\frac{L_1}{2} + \omega_B
\right]
\bigcup \Big[L_1 - \omega_B,L_1\Big]
\right)
\times [0,L_2] \times [0,L_3] \, ,
\label{omegab}
\\
\Omega_P
&=&
\left( 
\left[\omega_P,\frac{L_1}{2}-\omega_P\right] 
\bigcup \left[\frac{L_1}{2} + \omega_P,L_1 - \omega_P\right]
\right)
\times [0,L_2] \times [0,L_3] \, ,
\label{omegap}    
\end{eqnarray}
where $\omega_B$ and $\omega_P$ satisfy
\begin{equation}
0 < \omega_P \le \omega_B < \frac{L_1}{4} \, .
\label{standineq}
\end{equation}
The subdomain $\Omega_B$ contains the layers of permanent charges, where we have observed the large difference between the results obtained by the PNP system~(\ref{NernstPlanck})--(\ref{Poisson}) and BD~simulations. In particular, we will use BD in domain $\Omega_B$ to follow trajectories of individual ions, while we use macroscopic concentration profiles $c^\pm$ in domain $\Omega_P.$ In $\omega_P = \omega_B$, then the intersection of $\Omega_P$ and $\Omega_B$ only contains the two-dimensional boundaries of these regions, while we have an overlap region if $\omega_P < \omega_B$. We denote it by
\begin{equation}
O = \Omega_P \cap \Omega_B =
O_x \times [0,L_2] \times [0,L_3] \, ,
\label{overlapregion}
\end{equation}
where
$$
O_x
= 
\Big[\omega_P,\omega_B\Big] 
\bigcup 
\left[
\frac{L_1}{2} \! - \! \omega_B,\frac{L_1}{2} \! - \! \omega_P
\right]
\bigcup 
\left[
\frac{L_1}{2} \! + \! \omega_P,\frac{L_1}{2} \! + \! \omega_B
\right]
\bigcup
\Big[L_1 - \omega_B,L_1-\omega_P\Big].
$$
We use grid sizes~(\ref{gridsizes}) to discretize the PNP equations at meshpoints~(\ref{meshpoints}) and notation~(\ref{cijkphiijkdef}) to denote the values of concentrations $c^\pm$ at the meshpoints. The state of the multi-resolution system at time $t$ is given by 
\begin{eqnarray*}
&\mbox{(s1)} & \; \mbox{the values of concentrations} \; c^+_{i,j,k}(t) \; \mbox{and} \;
c^-_{i,j,k}(t) \; \mbox{at meshpoints} \; 
\mathbf{x}_{i,j,k} \in \Omega_P,
\\
&\mbox{(s2)} & \; \mbox{positions} \; \mathbf{X}_j^\pm(t)=[X_{j,1}^\pm(t),X_{j,2}^\pm(t), X_{j,3}^\pm(t)] \in \Omega_B, \; \mbox{for} \; j=1,2,\dots,N_B^\pm(t), 
\\
&\mbox{(s3)} & \; \mbox{potential} \; \phi_{i,j,k}(t) \; \mbox{for} \; i=1, 2, \dots, n_1, \; j=1, 2, \dots, n_2, \; k=1, 2, \dots, n_3,
\end{eqnarray*}
where $N_B^\pm(t)$ is the number of positive and negative ions described as individual particles by the BD approach. The number of simulated ions $N_B^\pm(t)$ depends on time~$t$. Since we have, in general, $N_B^{+}(t) \ne N_B^{-}(t)$, the BD subsystem does not satisfy electroneutrality on its own, but the system will be electroneutral when both parts of the system (s1) and (s2) are considered together. In particular, we will not use the Ewald summation, but the potential $\phi$ at meshpoints~(\ref{meshpoints}) will be calculated as the sum of two terms given by equation~(\ref{phiijkdec}), where potential $\phi^m_{i,j,k}$ corresponds to the mobile charges expressed either as concentrations $c^\pm$ of ions in $\Omega_P$ or as individual ions in $\Omega_B$. 

One iteration of the multi-resolution algorithm is given in Table~\ref{table4} as Algorithm [M1]--[M8]. Given the state of the multi-resolution system (s1)--(s3) at time $t$, Algorithm [M1]--[M8] calculates the state of the system (s1)--(s3) at time $t+\Delta t.$ In step~[M1], we evolve the concentrations $c^\pm$ from time $t$ to time $t+\Delta t$
by using~(\ref{eulera})--(\ref{eulerb}). Since the state of the multi-resolution system (s1)--(s3) only defines $c^\pm$ in subdomain $\Omega_P$, we first extend it by zero to the rest of the simulation domain by putting
\begin{equation}
c^\pm_{i,j,k}(t) := 0 \quad \mbox{for meshpoints} \; 
\mathbf{x}_{i,j,k} \in \Omega \setminus \Omega_P.
\label{zeroext}
\end{equation}
Then we can apply equations~(\ref{eulera})--(\ref{eulerb}) to calculate the time evolution of concentrations $c^\pm$ over one time step $[t,t+\Delta t]$. Since the calculated concentrations are further modified in step~[M6] to take into account the transfer of ions between $\Omega_B$ and $\Omega_P$, we will add an extra asterisk to $c^\pm$ and denote the output of calculations in step~[M1] as $c^{*,+}_{i,j,k}(t+\Delta t)$ and $c^{*,-}_{i,j,k}(t+\Delta t)$, rewriting equations (\ref{eulera})--(\ref{eulerb}) as
\begin{eqnarray}
c^{*,+}_{i,j,k}(t+\delta t)
&:=&
c^+_{i,j,k}(t)
\,
+
\, \delta t \,
{\mathcal A}_{i,j,k} \big(c^+(t),\phi(t)\big) \, .
\label{eulerastar}
\\
c^{*,-}_{i,j,k}(t+\delta t)
&:=&
c^-_{i,j,k}(t)
\,
+
\, \delta t \,
{\mathcal A}_{i,j,k} \big(c^-(t),\phi(t)\big) \, .
\label{eulerbstar}
\end{eqnarray}
Equations~(\ref{eulerastar})--(\ref{eulerbstar}) are formulated with time step $\delta t$, which we use in Section~\ref{sec43} to solve PNP equations. If $\delta t = \Delta t$, this means that step~[M1] is equivalent to step~[B1] from Algorithm~[B1]--[B4] for solving the PNP equations. However, we can also choose $\delta t < \Delta t$ such that $\Delta t/\delta t$ is an integer, and apply equations~(\ref{eulerastar})--(\ref{eulerbstar}) multiple times to calculate concentrations $c^{*,\pm}$ at time $t+\Delta t$ in step~[M1]. 

\begin{table}
\boxed{\hbox{\hskip 1mm\hsize=0.955\hsize\vbox{\vskip 0.5mm
\parindent -8.4mm \leftskip 8.4mm
[M1] Extend the values of concentrations $c^+_{i,j,k}(t)$ and $c^-_{i,j,k}(t)$ outside of the sub\-domain $\Omega_P$ by~(\ref{zeroext}) and calculate the concentrations $c^{*,+}_{i,j,k}(t+\Delta t)$ and $c^{*,-}_{i,j,k}(t+\Delta t)$ by equations~(\ref{eulerastar})--(\ref{eulerbstar}).
\par \vskip 1mm
[M2] Calculate drift terms ${\mathbf{a}}_j^\pm(t)$, for $j=1,2,\dots,N_B^\pm(t),$ using steps~[A1]--[A4] where $N$ is replaced by $N_B^\pm(t)$ and $\beta=\infty.$ Put \hfill\break
\rule{0pt}{3.7mm} \hskip 1cm ${\mathbf{a}}_j^\pm(t) := {\mathbf{a}}_j^\pm(t)
- \alpha^\pm q^\pm  \, {\overline{\nabla} \phi}\big({\textbf X}_j^\pm\big)$,
$\quad$for $j=1,2,\dots,N_B^\pm(t),$
\hfill\break
where ${\overline{\nabla} \phi}\big({\textbf X}_j^\pm\big)$ is a discretized gradient given by equation~(\ref{fmesh}).
\par \vskip 1mm
[M3]  Generate $6N_B(t)$ coordinates of vectors $\boldsymbol{\xi}_j$, for $j=1,2,\dots,N_B^\pm(t)$, as normally distributed numbers with zero mean and unit variance. Calculate the positions of mobile ions at time $t+\Delta t$ by using equation~(\ref{BDSDEform3DEMwithalpha}).
\par \vskip 1mm
[M4] Calculate $\mu^\pm(t+\Delta t)$ using \eqref{defalpha}. Put $\gamma^\pm = \gamma_{(i)}^\pm$ where $\gamma_{(i)}^\pm$ is given
by (\ref{defbetas}). 
\par \vskip 1mm
[M5] Generate two uniformly distributed random numbers $r^\pm$ in $(0,1)$. \hfill\break
If $r^\pm < \mu^\pm(t+\Delta t)$, then create new ion in $\Omega_B \setminus \Omega_P$ according to the probability density $p_B^\pm(x, t+\Delta t)$ defined in \eqref{eq:p22} and \hfill\break set $\gamma^\pm = \gamma_{(ii)}^\pm$ where $\gamma_{(ii)}^\pm$ is given by (\ref{defbetas}).
\par \vskip 1mm
[M6] Terminate trajectories of BD ions which landed in $\Omega_P \setminus \Omega_B$. \hfill\break
Calculate concentrations $c^{+}_{i,j,k}(t+\Delta t)$ and $c^{-}_{i,j,k}(t+\Delta t)$ in $\Omega_P$ by~(\ref{cupdate}).
\par \vskip 1mm
[M7] Calculate the discrete Fourier transform $f_{i^\prime,j^\prime,k^\prime}$ of \hfill\break
$\displaystyle
-\frac{1}{\varepsilon_0 \, \varepsilon} \left[
q^+ \, c^+_{i,j,k}
\,+\, 
q^- c^-_{i,j,k}
+
q^+
\sum_{j=1}^{N_B^+}
\frac{\delta({\mathbf X}_j^\pm)_{i,j,k}}{\Delta x_1 \,\Delta x_2 \,\Delta x_3}
+
q^-
\sum_{j=1}^{N_B^-}
\frac{\delta({\mathbf X}_j^-)_{i,j,k}}{\Delta x_1 \,\Delta x_2 \,\Delta x_3}
\right]
$
using the fast Fourier transform algorithm. 
\par \vskip 1mm
[M8] Calculate $\phi^m_{i,j,k}$ as the inverse discrete Fourier transform of
$
\lambda(i^\prime,j^\prime,k^\prime)   
f_{i^\prime,j^\prime,k^\prime}
$,
where $\lambda(i^\prime,j^\prime,k^\prime)$ is given by~(\ref{lambdacoef}), using the inverse fast Fourier transform algorithm.
Calculate $\phi_{i,j,k}(t+\Delta t)$ using equation~(\ref{phiijkdec}).}}}
\vskip 1mm
\caption{\label{table4}
One iteration of the multi-resolution algorithm.}
\end{table}

In step~[M2], we calculate drift terms ${\mathbf{a}}_j^\pm(t)$, for $j=1,2,\dots,N_B^\pm(t)$, which are used to update the positions of BD ions over one time step using equation~(\ref{BDSDEform3DEMwithalpha}). The calculation of drift terms follows steps~[A1]--[A4], where $N$ is substituted by $N_B^\pm(t)$ and we formally use $\beta=\infty$, {\it i.e.} we only consider the Lennard-Jones forces between all permanent and mobile ions in formulas used in~[A1]--[A4] and add their contributions to to drift terms ${\mathbf{a}}_j^\pm(t)$ in step~[M2]. Then we also add all electric contributions in step~[M2] by differentiating electric potential using the central difference scheme as
\begin{equation}
{\overline{\nabla} \phi}\big({\textbf x}\big)
=
\left[ 
\frac{\phi_{i+1,j,k}-\phi_{i-1,j,k}}{2\Delta x_1},
\frac{\phi_{i,j+1,k}-\phi_{i,j-1,k}}{2\Delta x_2},
\frac{\phi_{i,j,k+1}-\phi_{i,j,k-1}}{2\Delta x_3}
\right],
\label{fmesh}
\end{equation}
where
\begin{equation}
i = \left\lceil \frac{x_1}{\Delta x_1} \right\rceil,
\quad
j = \left\lceil \frac{x_2}{\Delta x_2} \right\rceil,
\quad
k = \left\lceil \frac{x_3}{\Delta x_3} \right\rceil,
\quad
\mbox{for}
\quad {\mathbf{x}} = [x_1,x_2,x_3],
\label{transformula}
\end{equation}
and $\lceil \cdot \rceil$ denotes the ceiling function, rounding a real number up to the nearest integer. Here, we again use the convention that indices $i,$ $j$ and $k$ are periodic with periods $n_1$, $n_2$ and $n_3$, respectively. Step~[M3] is the same as step [A6], where we perform the BD part of the simulation by substituting the calculated drift terms ${\mathbf{a}}_j^\pm(t)$, for $j=1,2,\dots,N_B^\pm(t)$, into equation~(\ref{BDSDEform3DEMwithalpha}). 

While concentration $c^\pm$ was equal to zero in $\Omega_B \setminus \Omega_P$ at time $t$ by our definition~(\ref{zeroext}), we will have a nonzero concentrations $c^{*,\pm}$ at time $t+\Delta t$. The mean number of ions `spilling over' to $\Omega_B \setminus \Omega_P$ during the time interval $[t, t+\Delta t]$ is calculated in step~[M4] by
\begin{equation}
\label{defalpha}
\mu^\pm(t + \Delta t) 
=
\int_{\Omega_B\setminus\Omega_P} 
c^{*,\pm}({\mathbf{x}}, t + \Delta t) \, \mbox{d}{\mathbf{x}}\,.
\end{equation} 
If $\Delta t$ is sufficiently small, then we have $\mu^\pm(t + \Delta t) \ll 1$. In particular, $\mu^\pm(t + \Delta t)$ can be interpreted as the probability of introducing new ion into $\Omega_B\setminus\Omega_P$. Its new position is sampled in step [M5] according to the probability distribution
\begin{equation}
\label{eq:p22}
p_B^\pm({\mathbf{x}}, t+\Delta t) =
\frac{c^{*,\pm}({\mathbf{x}}, t+\Delta t)}{\mu^\pm(t+\Delta t)}\,,
\qquad \mbox{for} \; {\mathbf{x}} \in \Omega_B \setminus \Omega_P \,.
\end{equation}
To ensure the conservation of the total number of particles in the whole domain $\Omega$, we also calculate scaling factors $\gamma^\pm$ in steps~[M4]--[M5], which are then used in step~[M6] to rescale the concentration profiles $c^{*,\pm}$ in $\Omega_P$. The scaling factors are given by
\begin{equation}
\gamma_{(i)}^\pm = \frac{N - N_B^\pm(t)}{N - N_B^\pm(t) - \mu^\pm(t + \Delta t)}\,,
\qquad
\gamma_{(ii)}^\pm = \frac{N - N_B^\pm(t)-1}{N - N_B^\pm(t) - \mu^\pm(t + \Delta t)}\,,
\label{defbetas}
\end{equation}
where $N_B^+(t)$ (resp. $N_B^-(t)$ is the number of individually simulated positive (resp. negative) ions in the BD~domain $\Omega_B$ at time $t$.

In step~[M6], we identify the ions which left the subdomain $\Omega_B$. They are removed from the simulation and added as Dirac delta function to obtain new concentration profiles in $\Omega_P$ at time $t+\Delta t$. More precisely, we define the discretized version of the Dirac delta function centered at position ${\mathbf x}$ by 
\begin{equation}
\delta({\mathbf x})_{i,j,k}
=
\begin{cases}
1
& \;\;\;\mbox{for $i$, $j$ and $k$ given by (\ref{transformula})};\\ 
0 & \;\;\;\mbox{otherwise.}
\end{cases}
\label{transdelta}
\end{equation}
Then the concentration update in step~[M6] can be written as
\begin{equation}
\;c^\pm_{i,j,k}(t+\Delta t)
\,=\,
\gamma^\pm
\,
c^{*,\pm}_{i,j,k}(t+\Delta t)
\;+
\!\!\!\sum_{\ell \in {\mathcal{J}}^\pm}
\frac{\delta({\mathbf X}_\ell^\pm(t+\Delta t))_{i,j,k}}{\Delta x_1 \,\Delta x_2 \,\Delta x_3}
\label{cupdate}
\end{equation}
where the set ${\mathcal{J}}^+$ (resp. ${\mathcal{J}}^-$) contains indices of all terminated positive (resp. negative) ions. In particular, we transform the position of each terminated positive (resp. negative) ion into its nearest mesh point $(i,j,k)$ by applying equation~(\ref{transformula}) and add $1/(\Delta x_1 \,\Delta x_2 \,\Delta x_3)$ to the corresponding concentration $c^+$ (resp. $c^-$) at this mesh point.  The concentration update~(\ref{cupdate}) preserves the number of particles, which we can formulate using our state variables (s1)--(s3) as
\begin{equation}
N
=
N_B^\pm(t)
+
\int_{\Omega_P} 
c^{\pm}({\mathbf{x}}, t) \, \mbox{d}{\mathbf{x}}
=
N_B^\pm(t)
+
\sum_{i=1}^{n_1}
\sum_{j=1}^{n_2}
\sum_{k=1}^{n_3}
c^\pm_{i,j,k}(t)
\,
\Delta x_1 \,\Delta x_2 \,\Delta x_3\,,
\label{conservationofparticles}
\end{equation}
where we have used the extension~(\ref{zeroext}) on the right-hand side, {\it i.e.} the summation on the right-hand side can be restricted to the mesh points, which are in $\Omega_P.$ 

Finally, steps [M7]--[M8] calculate the potential $\phi_{i,j,k}(t+\Delta t)$ using a similar approach as in steps [B2]--[B4]. In step~[M7], the right-hand side of the discretized Poisson equation~(\ref{potcdisc}) is modified to take into account all mobile ions at time $t+\Delta t$. In particular, potential $\phi_{i,j,k}^m$ (calculated in step~[M8] by the inverse fast Fourier transform algorithm) will contain electric potential corresponding to all mobile ions, which are either represented as individual ions or as concentration fields. In step~[M8], we add potential $\widehat{\phi}_{\mathrm{p}}\big(\mathbf{x}\big)$
corresponding to the permanent charges using equation~(\ref{phiijkdec}) at time $t+\Delta t$. As it has been the case of our BD~simulations in Section~\ref{sec42} or our PNP solutions, the permanent potential can be precomputed at the beginning of the simulation by solving equation~(\ref{potentialpermanent}).

\begin{figure}[t]
\epsfig{file=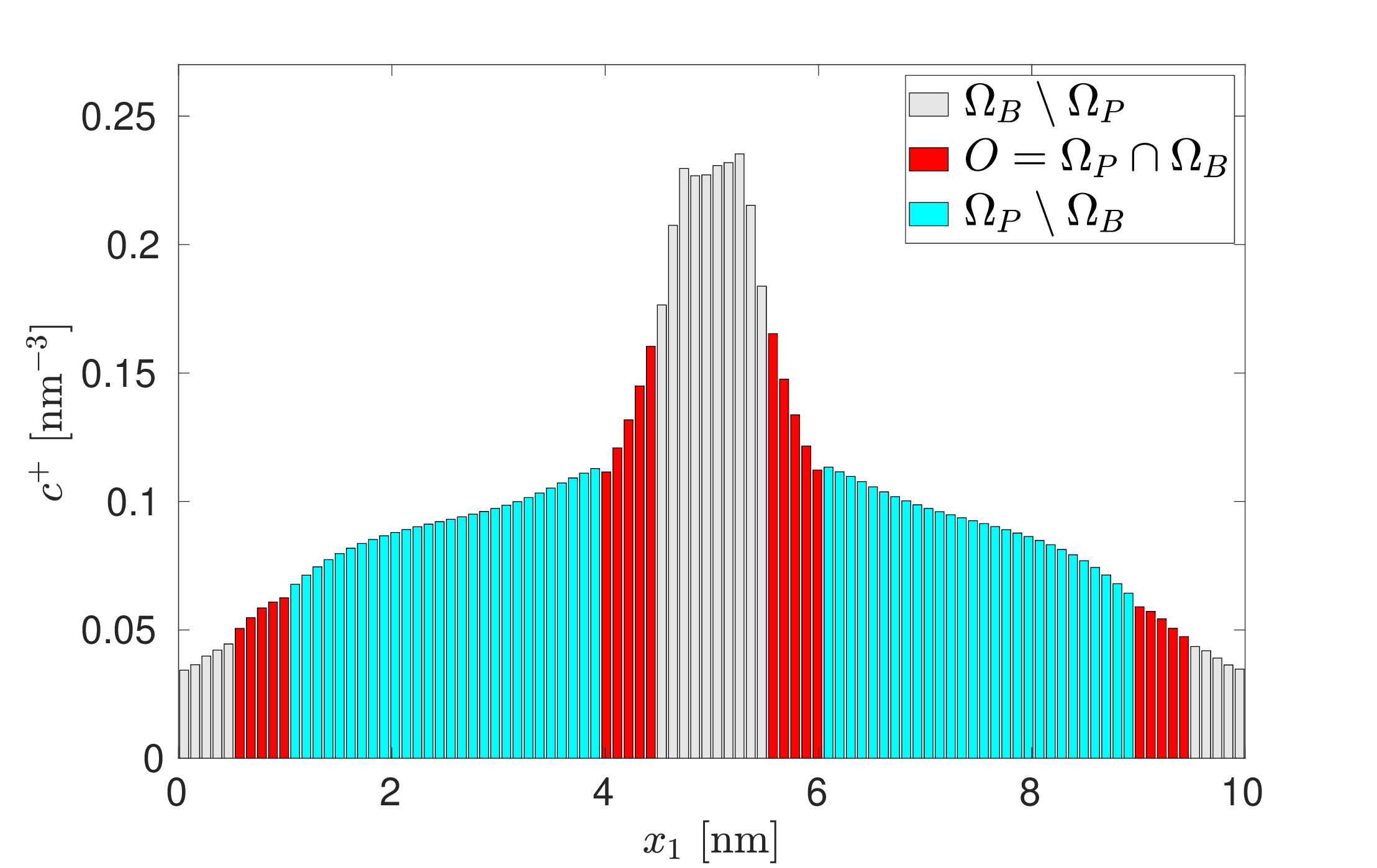,height=3.94cm}
\hskip 0.1mm 
\epsfig{file=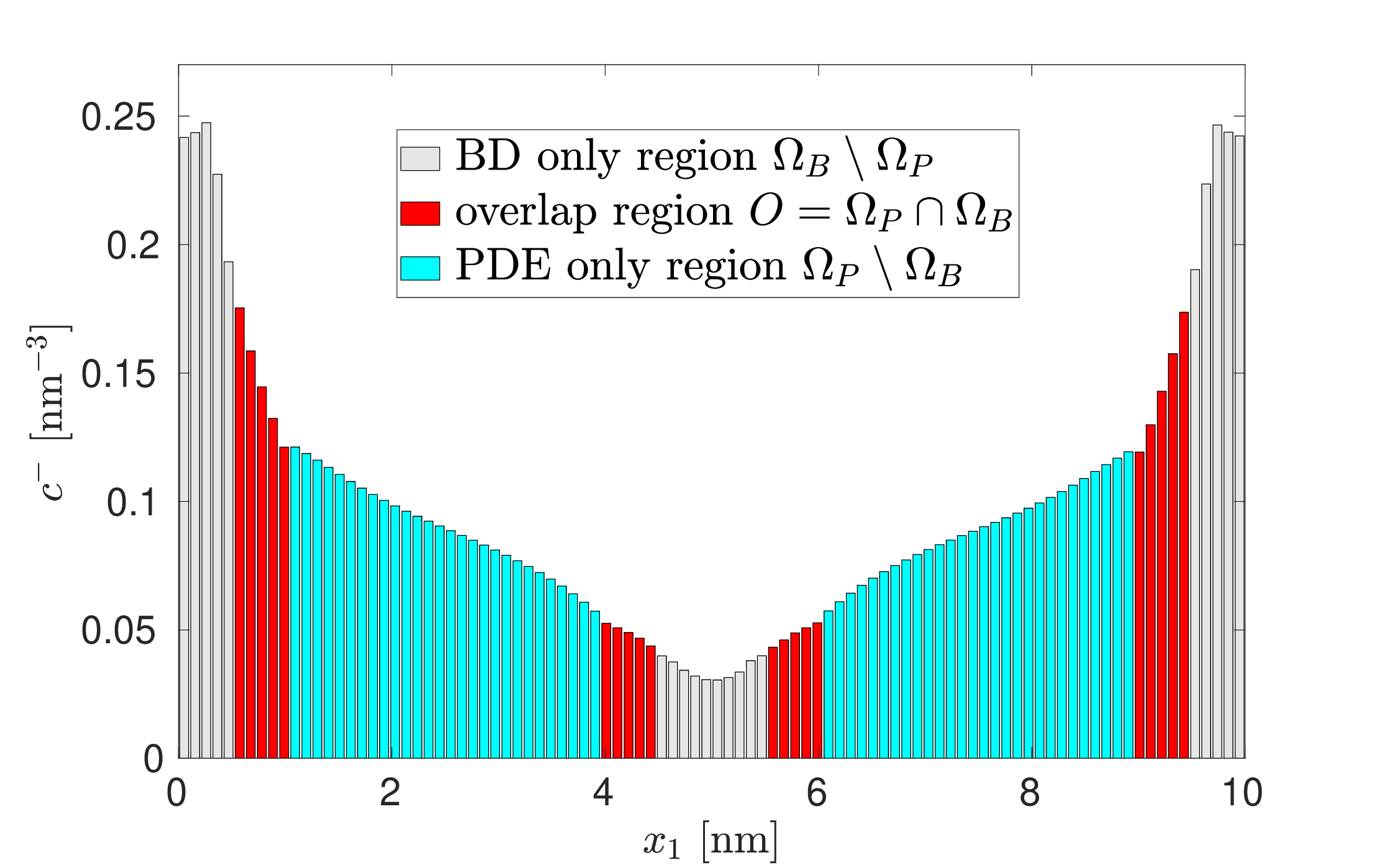,height=3.94cm}
\vskip -4cm
(a) \hskip 5.8cm (b)
\vskip 3.45cm
\caption{The equilibrium distributions calculated using Algorithm~{\rm [M1]--[M8]} for $N=100$, $N_p=6$, $n_1 = n_2 = n_3 = 96$, $L_1 = L_2 = L_3 = 10 \, \mbox{{\rm nm}}$, $\omega_P = 5 L_1/n_1$ and $\omega_B = 10 L_1/n_1$ for \hfill\break {\rm (a)} mobile {\rm Na}$^+$ ions;  and {\rm (b)} mobile Cl$^-$ ions.}
\label{fig3}
\end{figure}%

In Figure~\ref{fig3}, we present results calculated using Algorithm~[M1]--[M8] applied to our illustrative model with $N=100$ mobile Na$^+$ ions and $N=100$ mobile Cl$^+$ ions in domain $\Omega$ given by~(\ref{cuboidomega}), where $L_1 = L_2 = L_3 = 10 \, \mbox{nm}$. The permanent charges are at locations~(\ref{permNachargeslocations})--(\ref{permClchargeslocations}) with $N_{\mathrm{p}}=6$. We use $n_1 = n_2 = n_3 = 96 = 3 \times 2^5$ mesh points in each direction to discretize the Poisson equation, {\it i.e.} our spatial resolution for the calculation of electric potential is the same as we used in Section~\ref{sec43}. The mesh sizes are given by~(\ref{gridsizes}), and we use $\omega_P = 5 \Delta x_1$ and $\omega_B = 10 \Delta x_1$. In Figure~\ref{fig3}, we present the calculated equilibrium densities as functions of the first coordinate, $x_1$, integrating over the $x_2$ and $x_3$ coordinates. To highlight the multi-resolution nature of the simulation, the concentrations $c^\pm$ in the region $\Omega_P \setminus \Omega_B$ are denoted using light blue (cyan) bars, the histograms in the region $\Omega_B \setminus \Omega_B$  are visualized using gray bars and the red bars show the average concentration in the overlap region~(\ref{overlapregion}), where some mobile ions are treated as concentration profiles and some mobile ions are described as individual particles (we add their contributions together to obtain the red bars).
Since the multi-resolution simulation explicitly includes the Lennard-Jones potentials around permanent charges, the density $c^\pm$ does not accumulate in the mesh points close to the permanent charges, which happens for the PNP solution in Figure~\ref{fig2}.

\section{Discussion}

\label{sec6}

In this paper, we have derived the PNP system~(\ref{NernstPlanck})--(\ref{Poisson}) from the BD model~(\ref{BDSDEform}) in Section~\ref{sec3} in the limit $N \to \infty$, where $N$ is the number of simulated ions. While the PNP system~(\ref{NernstPlanck})--(\ref{Poisson}) is justified in a bulk solution~\cite{Schuss:2001:DPN}, the error term~(\ref{errorforce}) can have significant effects on the system dynamics close to the biological structures of interest. For example, considering applications to modelling ion channels, the PNP system~(\ref{NernstPlanck})--(\ref{Poisson}) can be used in regions
far away from an ion channel, but its applicability to the passage
of ions through the ion channel is questionable, especially in its narrowest part as indicated by computational studies in the literature~\cite{Corry:2000:TCT,Song:2011:TAN,Im:2002:IPS}. In Section~\ref{sec4}, we have presented an illustrative system with layers of permanent charges showing that the Lennard-Jones potential creates exclusion zones around permanent charges which are not captured by the PNP system~(\ref{NernstPlanck})--(\ref{Poisson}). Such observations motivate the development of a multi-resolution approach in Section~\ref{sec5}, which uses PDEs~(\ref{NernstPlanck})--(\ref{Poisson})
in the bulk and more detailed BD simulations in the regions close to the permanent charges. The modeling accuracy in regions described by BD could be further enhanced by incorporating a molecular dynamics description~\cite{Im:2002:IPS,ErbanTogashi}.

To design the multi-resolution scheme in Section~\ref{sec5}, we have extended the PDE-assisted BD approach for reaction-diffusion systems, developed in~\cite{Franz:2013:MRA}, to models of charged particles. The resulting multi-resolution Algorithm~[M1]--[M8] uses the discrete Fourier transform to solve the Poisson equation. In particular, we need to transform positions of individual ions to the corresponding mesh points by~(\ref{transdelta}) and use the calculated mesh-based potential to extract the forces by~(\ref{fmesh}).
Equations~(\ref{fmesh}) and~(\ref{transdelta})
possess the necessary symmetry to ensure that each individual ion does not respond to the potential arising from its own charge. However, this remains a different approach compared to the BD method that employs the Ewald summation. In the case of reaction-diffusion systems~\cite{Franz:2013:MRA}, the multi-resolution algorithm used a BD approach without modifications, because there were no long-range forces to consider. If a modeller wants to use the Ewald summation for interactions between individually simulated particles in $\Omega_B$, then they would need to ensure the electroneutrality in the BD subdomain, required by the Ewald summation. To further improve the accuracy, there is a potential to use some hybrid approaches like particle-particle-particle mesh algorithms~\cite{Hockney:1988:CSU,Gibbon:2002:LRI}, or to use multi-grid approaches~\cite{multigrid}.

In our implementation of Algorithm~[M1]--[M8] in Section~\ref{sec5}, the multi-resolution method includes the overlap region~(\ref{overlapregion}), where both the PDE description and BD exist in parallel. In some multi-resolution approaches, it is possible to couple models with different resolutions using a simple interface, as it has been shown for coupling BD simulations with a coarser description given by a compartment-based model (reaction-diffusion master equation) in~\cite{FleggChapman}. The advantage of a simple interface over an overlap region is that this can simplify some aspects of the software implementation of the resulting multi-resolution method~\cite{Robinson:2015:MRD}. 
It remains an open question for future investigation whether the BD modelling of charged particles can be coarse-grained as a compartment-based stochastic model which could potentially avoid using an overlap region in a multi-resolution scheme. However, if a BD model is coarse-grained using mean-field PDEs (like the PNP system in this paper), then the use of an overlap region improves the accuracy, as it was previously shown for reaction-diffusion systems in the literature~\cite{Franz:2013:MRA,Smith:2018:ARM}. The~benefits of overlap (bridging, blending, transition) regions have been also demonstrated in other multi-resolution approaches, including coupling molecular dynamics and BD simulations~\cite{Erban:2014:MDB,Erban:2016:CAM,delRazo2021}, compartment-based models with macroscopic PDEs~\cite{KangErban,Yates2015,Smith2018}, to bridge particle-based BD models with macroscopic PDEs~\cite{Yates2020,delRazo2025}, to coarse-grain the chemical master equation in the state space~\cite{Duncan2016}, and more broadly to connect atomistic and continuum modelling in material science~\cite{Miller2009,Luskin,
vanKoten2011,Parks2008,Xiao2004,Smith2024}.

Considering our illustrative example in Section~\ref{sec5}, the domain decomposition into the BD and PDE subdomains $\Omega_B$ and $\Omega_P$ remained fixed throughout the entire simulation. In some application areas, it is necessary to consider extending multi-resolution techniques to scenarios where the BD subdomain $\Omega_B$ changes over time~\cite{Robinson:2014:ATM}, or the domain undergoes temporal growth~\cite{Baker2010,Smith2021}.

\vskip 7mm

\begingroup
\renewcommand{\section}[2]{}%

\endgroup

\end{document}